\documentclass[]{ar2e}

\usepackage[oztex]{graphicx}

\begin{document}

\jname{{\it Annu. Rev. Phys. Chem.\/}}
\jyear{2002}
\jvol{52}

\title{Molecular Theory of Hydrophobic Effects: ``She is too mean to
have her name repeated.''\footnote{W. Shakespeare: {\em All's well that ends well}}}

\markboth{Lawrence  R.  Pratt}{Hydrophobic Effects}

\author{Lawrence  R.  Pratt
\affiliation{Theoretical Division, Los Alamos National Laboratory, Los Alamos, NM 87545, USA,
phone:  505-667-8624, fax:  505-665-3909, email:  lrp@lanl.gov}}


\begin{abstract}
This paper reviews the molecular theory of hydrophobic effects relevant
to biomolecular structure and assembly in aqueous solution.  Recent
progress has  resulted in simple,  validated molecular statistical
thermodynamic theories and clarification of confusing theories of
decades ago.  Current work is resolving effects of wider variations of
thermodynamic state, {\it e.g.\/} pressure denaturation of soluble
proteins, and more exotic questions such as effects of surface chemistry
in treating stability of macromolecular structures in aqueous solution
\end{abstract}
\vfill
\pagebreak
\maketitle

\section*{INTRODUCTION}\addcontentsline{toc}{section}{INTRODUCTION}

In the past several years there has been a breakthrough,
associated with the efforts of a theoretical collaboration at Los
Alamos\cite{Hummer:PNAS:96,Garde:PRL:96,Pratt:ECC,Hummer:PNAS:98,%
Hummer:PRL:98,Hummer:JPCB:98,Pohorille:PJC:98,Gomez:99,Garde:99,%
Pratt:NATO99,HummerG:Newphe} but with important
antecedents\cite{Lee:85,Pohorille:JACS:90,%
Pratt:PNAS:92,Palma,Chandler:PRE:93},  on the problem of the molecular
theory of hydrophobic effects.  That breakthrough is the justification
for this review.

One unanticipated consequence of that work has been the clarification of
the `Pratt-Chandler theory'\cite{Pratt:JCP:77}. Judged empirically, that
theory was not less successful than is typical of molecular theories of
liquids.  But the supporting theoretical arguments had never been
compelling and engendered significant confusion. That confusion was
signaled already in 1979 by the view\cite{Chan:79}: ``The reason for the
success of their theory may well be profound, but could be accidental. 
We cannot be sure which.'' Today the correct answer is `both,' though
accidental first.  In an amended form, it is a compelling theory.  In
reviewing these developments, a significant volume of intervening
theoretical work must eventually be viewed again in this new light.  In
addition, the work that clarified the Pratt-Chandler theory suggested
several improvements and extensions, and was deepened by the parallel 
development of the molecular quasi-chemical theory of
solutions\cite{PrattLR:Quatal,Pratt:ES:99,Rempe:JACS:2000,%
Rempe:FPE:2001,HummerG:Newphe,Pratt:2001}.  On this basis, I predict an
extended period of consolidation  of the theory of these systems and
inclusion of more realistic, interesting, and exotic instances.

An attitude of this review is to respect a scholarly patience in
addressing the foundations without prejudging the speculations ellicited
by the fascinating biophysical motivations.  Thus, physical chemists
working at strengthening those foundations are the audience for this
review.

Nevertheless, clarity of the biophysical goals is important.  Thus, an
uncluttered expression of the motivation and the basic problem is
essential. The molecular theory of hydrophobic effects is an unsolved
facet of a molecular problem foundational to biophysics and
biochemistry: the quantitative molecular scale understanding of the
forces responsible for structure, stability, and function of
biomolecules and biomolecular aggregates. I estimate that the term
`hydrophobic' appears in every biophysics and biochemistry textbook.  An
intuitive definition of hydrophobic effects is typically assumed at the
outset. Hydrophobic effects are associated with demixing under standard
conditions of oil-like materials from aqueous solutions. A more refined
appreciation of hydrophobic effects acknowledges that they are a part of
a subtle mixture of interactions that stabilize biomolecular structures
in aqueous solution over a significant temperature range while
permitting sufficient flexibility for the biological function of those
structures.
	
Preeminent characteristics of these hydrophobic interactions are
temperature dependences, and concurrent entropies, that can be
exemplified by cold denaturation of soluble proteins\cite{Privalov:90}.
If hydrophobic effects stabilize folded protein structures, then folding
upon {\em heating} suggests that hydrophobic
interactions become {\em stronger} with increasing temperature through
this low temperature regime. This is a counter-intuitive observation.
	
A primitive correct step in relieving this contrary intuition is the
recognition that water molecules of the solution participate in this
folding process\cite{ParsegianVA:Osmscp}.  Specific participation by
small numbers of water molecules is not unexpected but also isn't the
mark of hydrophobic effects. For hydrophobic effects, on the contrary, a
large collection of water molecules are involved nonspecifically. It is
the statistics of the configurations of these water molecules at the
specified temperature that lead to the fascinating entropy issues.
Because of the significance of these entropies, hydrophobic effects are
naturally a topic for molecular statistical thermodynamics.

Given the acknowledged significance of this topic, it is understandable
that the literature that appeals to them is vast. Many researchers from
a wide range of backgrounds and with a wide variety  of goals have
worked on these problems.   Thus, the `unsolved' assertion  will
challenge those researchers. But the `unsolved' assertion also reflects
a lack of integration of principles, tools, and results to form a
generally accepted mechanism of hydrophobic effects.  For example, it is
widely, but not universally, agreed that the hydrogen bonding
interactions between water molecules are a key to understanding
hydrophobic effects. Conventional molecular simulation calculations with
widely accepted molecular interaction models for small hydrophobic
species in water broadly agree  with experimental results on such
systems\cite{DASHEVSKYVG:SOLHIN,OWICKIJC:MONCIE,SWAMINATHANS:MONSSD,%
GEIGERA:MOLSTH,PANGALIC:HYDHAP,BIGOTB:SAMMMS,POSTMAJPM:THECFW,%
OKAZAKIS:TEMETH,KINCAIDRH:ACCCMS,%
ROSSKYPJ:MOLLSO,RAPAPORTDC:HYDISM,TANIA:NONSWP,SWOPEWC:MOLMCS,Remerie:MP:84,%
LINSEP:MONSDA,JORGENSENWL:MONSAW,STRAATSMATP:FREHHM,ZICHIDA:SOLMRH,ZICHIDA:MOLCEL,FOISES:MONSAN,%
TANAKAH:INTMSH,FLEISCHMANSH:THEASS,KOOPOY:MODSHS,JORGENSENWL:FRETWF,RAOBG:HYDHFP,%
LINSEP:MOLSDA,GUILLOTB:COMSHH,%
LAAKSONENA:MOLNMW,CUMMINGSPT:SIMSWS,TANAKAH:HYDHIT,andaloro,FLEISCHMANSH:FRESMS,%
Lazaridis:92,Wallqvist:JCP:92,SUNYX:SIMTSF,LAZARIDIST:ENTHHN,Guillot:JCP:93,%
GIIOB:AQUNGC,SMITHDE:FREEIE,VANBELLED:MOLSMH,GUILLOTB:TEMTSN,ZENGJ:ENTHPM,SKIPPERNT:COMMWS,%
Beglov:94,FORSMANJ:MONSHI,MADANB:ROLHHF,MATUBAYASIN:MATWGH,%
MATUBAYASIN:THETHS:1,BUSHUEVYG:STRCHS,LAZARIDIST:SIMSTH,KAMINSKIG:FREHPL,%
headgordont:effsms,BEUTLERTC:FRECFS,Head-Gordon:JACS:95,SUNYX:HYDSMN,Wallqvist:JPC:95:a,Head-Gordon:PNAS:95,%
MANCERARL:TEMETH,AshbaughHS:EnthhE,Garde:PRE:96,Prevost:JPC:96,ChauPL:Curehs,%
ReM:Strctd,MatubayasiN:Theths:2,DurellSR:Atoats,WallqvistA:theoth,SkipperNT:Comsst,%
HaymetADJ:Hydrmt,GardeS:HydiCe,LinCL:Pretfe,%
ManceraRL:Hydbth,MengEC:Moldst,LyndenBellRM:hydhbs,FlorisFM:Freeei,DeJongPHK:Hydhm,%
RadmerRJ:Freecm,mancerarl:aggmas,Silverstein:98a,SilversteinKAT:simmwh,IkeguchiM:Rolhbh,%
ArthurJW:Solnsw,PanhuisMIH:moldsc,MountainRD:Hydsh,ErringtonJR:Molspe,%
mancerarl:comste,Silverstein:99,TomasOliveiraI:Thecfw,Arthur:JCP:99,PomesR:Calecp,%
UrahataS:MonCst,FoisE:Hydecs,SlusherJT:Acceic,SomasundaramT:pasgtl,ChauPL:Comsts,%
SmithPE:Comsce,MadanB:Chawsi,GuisoniN:Squwas,SvishchevIM:Solstd,RasaiahJC:Strast,%
UrbicT:twomwT,NoworytaJP:Dynasi,SchurhammerR:ArehA+,SouthallNT:mechsd,GallicchioE:Entcda,%
Hernandez-CobosJ:hydhma,BergmanDL:Isheuw,RaschkeTM:Quathi,GardeS:Temdhh,%
KairaA:Salshs}.  In this sense, everything is known.  Nevertheless, this
complete knowledge hasn't achieved a consensus for a primitive mechanism
of hydrophobic effects.  By `mechanism' we mean here a simpler, physical
description that ties together otherwise disparate observations. Though
this concept of mechanism is less than the complete knowledge of
simulation calculations, it is not the extreme  `poetic ``explanation'''
famously noted by Stillinger\cite{Stillinger:73}.  The elementary
simplifications that lead to a mechanism must be more than
rationalizations; they must be verifiable and consistent at a more basic
level of theory, calculation, and observation.

The `breakthrough' mentioned in the first paragraph is particularly
exciting because it hints at such a mechanism for the most
primitive hydrophobic effects.  A great deal more reseach is called for,
certainly.  But discussion of that development is a principal feature of this
review.

\section*{WHAT CHANGED}\addcontentsline{toc}{section}{WHAT CHANGED}

The breakthrough required a couple steps.  The first step was the
realization that feasible statistical investigations of spontaneous
formation of atomic sized cavities in liquid solvents should shed light on
operating theories of hydrophobic
hydration\cite{Lee:85,Pohorille:JACS:90,Pratt:91,Pratt:PNAS:92,Palma}. Those
studies could be based upon the formal
truth\cite{Pratt:ECC,Pratt:NATO99} 
\begin{eqnarray}
\Delta\mu_\mathrm{A} = -RT \ln p_0 
\label{p0}
\end{eqnarray} 
where $\Delta\mu_\mathrm{A}$ is the interaction contribution to the
chemical potential of a hard core hydrophobic solute of type A, and
$p_0$  the probability that an observation volume defined by the
excluded volume interactions of A with water molecules would have zero
(0) occupants.  These cavity formation studies were not an attempt to
calculate hydration free energies for realistic hydrophobic solutes. 
The goal was just to examine simple theories and to learn how
different solvents might be distinguished on this basis.

\subsection*{The `small size'
hypothesis}\addcontentsline{toc}{subsection}{The `small size'
hypothesis}
There was also a significant physical idea alive at the time those
studies were undertaken\cite{Lee:85}:  ``The low solubility of nonpolar
solutes in water arises not from the fact that water molecules can form
hydrogen bonds, but rather from the fact that they are small in size.''
As a simple clear hypothesis, this view contributed to the breakthrough 
although the hypothesis was eventually disputed\cite{Pohorille:JACS:90,TangKES:ExcvsS}.

Packing and molecular sizes are important concerns for
liquids because they are dense.  The idea was that since water
molecules are smaller than, say, CCl$_4$ molecules, the
`interstitial' spaces available in liquid water would be smaller than
those in liquid CCl$_4$.  The first disputed point was that this
hypothesis was suggested by the approximate scaled particle
model\cite{Pierotti:76,LUCASM:SIZETN} and that model was known to have
flaws\cite{Ben-Naim:67,Stillinger:73} as applied to hydrophobic
hydration.  The second disputed point was that treatment of the
coexisting organic phase, liquid CCl$_4$ in this discussion, was less
convincing than the treatment of liquid water: the modeling of a CCl$_4$
molecule as a simple ball is an oft-convenient canard but shouldn't be
taken too literally.   Another significant consideration is that liquid
water is less dense on a packing fraction basis than most coexisting
organic solvents.

Eventually\cite{Pohorille:JACS:90,Pratt:91,Pratt:PNAS:92,Palma}, the
direct investigations at the low pressures of first interest indicated
that the probable spontaneously occurring cavities might be smaller in
typical organic solvents than in water, though the differences are
small.  What was decidedly different between water and organic solvents
was the flexibility of the medium to open cavities larger than the most
probable size. Water is {\em less} flexible in this regard, stiffer on a
molecular scale, than typical organic liquids. Fig.~1 gives a
macroscopic experimental perspective on this relative stiffness.
Furthermore, the results for the organic phase were not at all similar
to what scaled particle models suggested\cite{Pratt:PNAS:92,Palma}.  So
as an explanation of the distinction between water and common organic
solvents, the  `small size' mechanism must be
discounted\cite{Madan:BPC:94,
WallqvistA:theoth,Silverstein:JCE:98,Lazaridis:2001}. Nevertheless, we
can anticipate subsequent discussion by noting that the equation of
state of the solvent is important in establishing thermodynamic
signatures of hydrophobic hydration.

\subsection*{Transient cavities probing packing and
fluctuations}\addcontentsline{toc}{subsection}{Transient cavities probing packing and
fluctuations}
As an analysis tool for assessment of packing in disordered phases,
studies of cavity statistics should be more widely
helpful\cite{Postma:FSCS:82,Remerie:MP:84,TANAKAH:INTMSH,POHORILLEA:MOLAI,Wolfenden:94,%
BEUTLERTC:FRECFS,Kocher:Structure:96,Prevost:JPC:96,Crooks:PRE:97,FlorisFM:Freeei,%
Stamatopoulou:JCP:98,TanakaH:Cavdlw,TomasOliveiraI:Thecfw,%
IkeguchiM:Rolhbh,MountainRD:Voicew,ReM:Strctd,GardeS:Micdfs,%
intVeldPJI:Liqsvc,GardeS:Temdhh,Pratt:2001,KussellE:Excvps}.  The interesting work of
Kocher, {\em et al.}, \cite{Kocher:Structure:96} is notable.
Those calculations studied the cavity formation work in protein
interiors and that cavity formation work was seen to be larger than for
comparable organic solvents. In that respect, the packing of those
protein interiors was tighter, less flexible, than a simple oil droplet.   This
conclusion seems significant for our pictures of protein structures and
deserves further investigation.  Structural and compositional
heterogeneity are undoubtedly also important features of the cores of
globular proteins.  These observations should be helpful in distinguishing
the interiors of micelles from the cores of folded globular proteins\cite{LesemannM:Predtc}.

\subsection*{Modeling occupancy
probabilities}\addcontentsline{toc}{subsection}{Modeling occupancy
probabilities}

The decisive second step in achieving the present breakthrough was the
modeling of the distribution $p_n$ of which $p_0$ is the n=0
member\cite{Hummer:PNAS:96}. Several specific distributions $p_n$ had
been tried in analyzing the results of Pratt and
Pohorille\cite{Pratt:PNAS:92}. But it was eventually recognized that a
less specific approach, utilizing a maximum entropy procedure to
incorporate successively more empirical moment information, was prudent
and effective.  Surprisingly, direct determination of the distribution
$p_n$ showed that a two moment model
\begin{eqnarray}
- \ln p_n \approx \zeta_0 + \zeta_1 n + \zeta_2 n^2
\label{gaussian}
\end{eqnarray}
was accurately born-out in circumstances of computer simulation of
liquid water\cite{Hummer:PNAS:96}.  The parameters $\zeta_j$ are
evaluated by fitting of the predicted moments $\langle n^j \rangle$, j =
0, 1, $\ldots$ to  moment data. A practical virtue of
this two moment model is that the required moment data can be obtained
from long-available experimental results. A further surprise was that it
had previously been shown, in different contexts and with additional
assumptions\cite{Chandler:PRE:93,Percus:JP:93}, that theories of a
Percus-Yevick analog type had a structure derivable from a Gaussian or
harmonic density field theory.  The Pratt-Chandler theory was of this
Percus-Yevick analog type.  To the extent that the empirical observation
Eq.~\ref{gaussian} suggests a normal distribution, the Pratt-Chandler
theory is given a better foundation than was available at its
genesis\cite{Pratt:NATO99}.

\subsection*{Non-equivalence with Pratt-Chandler
theory}\addcontentsline{toc}{subsection}{Non-equivalence with
Pratt-Chandler theory}
In fact, the two moment model Eq.~\ref{gaussian} is {\em not} precisely the
same as the Pratt-Chandler theory.  There are several related direct
observations that can make that point clear. For example, the
probability model Eq.~\ref{gaussian} assigns probability weight only to
non-negative integer occupancies.  That is not the case for the harmonic
density field theory.   The restriction that such
density field theories should not permit the negative occupancy of any
subvolume is an  obvious but interesting requirement.  Additionally, the
Percus-Yevick theory for hard sphere mixtures can predict {\em
negative} probabilities\cite{MITCHELLDJ:HARSEC}. These are
technical issues, however, and the performance of the two moment model
Eq.~\ref{gaussian} gives strong and unexpected support for the
Pratt-Chandler theory.

There is a different respect in which the correspondence of the two
moment model Eq.~\ref{gaussian} with the Pratt-Chandler theory is
imprecise. The kinship indicated above is based upon calculation of
hydration free energies when the solvent can be idealized as a harmonic
density field.  That can be straightforwardly carried over to
consideration of nonspherical solutes.  For example, classic potentials
of mean force might be addressed by consideration of a diatomic solute
of varying bond length.  The Pratt-Chandler theory does not do that
directly but utilizes the structure of the Ornstein-Zernike equations
together with yet another Percus-Yevick style closure approximation.
Those distinctions have not yet been discussed fully.

\subsection*{`Good theories are either Gaussian or
everything'}\addcontentsline{toc}{subsection}{`Good theories are either
Gaussian or everything'} A curious feature of the $\ln p_n$ moment
modeling is that convergence of predictions for $\Delta\mu_\mathrm{A} =
-RT \ln p_0 $ with increasing numbers of utilized moments is
non-monotonic\cite{Pratt:NATO99,Hummer:JPCB:98,Gomez:99}.  The predicted
thermodynamic results are surprisingly accurate when two moments are
used but become worse with three moments before eventually returning to
an accurate prediction with many more moments available.  The two moment
model, and also the Pratt-Chandler theory, is fortuitous in this sense.
But this does conform to the adage that `good theories are either
Gaussian or everything.'

\section*{TECHNICAL OBSERVATIONS AND
EXTENSIONS}\addcontentsline{toc}{section}{TECHNICAL OBSERVATIONS AND
EXTENSIONS}

The theory above has always been understood at a more basic level than
the description above\cite{Pratt:NATO99}.  The most important
observation is that the Mayer-Montroll
series\cite{Stell:85,Pratt:NATO99} can be made significantly
constructive with the help of simulation
data\cite{Pohorille:JACS:90,Pratt:91,Pratt:NATO99} and those approaches
can be more physical than stock integral equation approximations.
Simulation data can provide successive terms in a Mayer-Montroll series.
 In that case the binomial moments $\langle {n \choose j} \rangle_0$ are
the stylistically preferred data\cite{Pratt:NATO99}.   Then the maximum
entropy modeling is a device for a resummation based upon a finite
number of initial terms of that series\cite{Pratt:NATO99}.  Several
additional technical points are helpful at this level.

\subsection*{Default Models}\addcontentsline{toc}{subsection}{Default models}

The adage that `good theories are either Gaussian or everything' is
serious but doesn't address the physical reasons why these
distributions are the way they are.  In fact, painstaking addition of
successive moments is not only painful but often unsatisfying.  And the
two moment model Eq.~\ref{gaussian} has been less satisfactory for every
additional case examined carefully beyond the initial one that was
connected with this breakthrough\cite{Hummer:PNAS:96}; recent examples
can be seen in Ref.~\cite{Pratt:2001,GardeS:Temdhh}.  It is better to
consider simple physical models for the distribution $p_n$ and
approximations
\begin{eqnarray}
- \ln \left[ { p_n \over {\hat p}_n } \right] \approx  \zeta_0 + \zeta_1 n + \zeta_2 n^2 + \ldots
\label{dmodel}
\end{eqnarray}
where  ${\hat p}_n$ is a model distribution chosen on the basis of
extraneous considerations. Since $p_n ={\hat p}_n$ in the absence
of further information, ${\hat p}_n$ is called the default model.
Utilization  of a default model in this way compromises the goal of predicting the
distribution and instead relies on the moments to adapt the default
model to the conditions of interest.   The default model of
first interest\cite{Hummer:PNAS:96} is ${\hat p}_n \propto 1/n!$  This
default model produces the uncorrelated result (the Poisson
distribution)  when the only moment used is $\langle n \rangle$. 
Another way to identify default models is to use probabilities obtained
for some other system having something in common with the aqueous solution
of interest\cite{Gomez:99}. 

An important  practical point is that this approach works better
when the default model is not too specific\cite{Rempe:FPE:2001}. This
can be understood as follows:  The moment information used
to adapt the default model to the case of interest is not particularly
specific.   If the default model makes specific errors, a limited amount
of that nonspecific data will not correct those errors adequately. 
This argument gives a partial rationalization for the accurate
performance of the flat default model that leads to Eq.~\ref{gaussian}.

In Eq.~\ref{gaussian} $\zeta_0 = \Delta\mu_\mathrm{A}/RT$ but it
is helpful to notice that this thermodynamic quantity can be
alternatively expressed as\cite{Pratt:NATO99}
\begin{eqnarray}
\Delta\mu_\mathrm{A} = RT\ln \left\{1 +  \sum_{n=1}\left({\hat{p}_n \over \hat{p}_0}\right) \exp
\left[-\sum_{k=1}^{k_{max}} \zeta_k {n \choose k} \right] \right\}
\label{eq:pf}
\end{eqnarray} 
where binomial moments through order $k_{max}$  are assumed and the
default model is included.  $\zeta_0$ does not appear on the right since
that normalization factor is being expressed through the thermodynamic
property.  The point is that this is a conventional form
of a partition function sum. The interactions are n-function interactions, in
contrast to density or $\rho$-functional theories,  but with strength parameters
adjusted to conform to the data available.  The fact that
$\Delta\mu_\mathrm{A}$ is extracted from a fully considered probability
distribution, and this consequent structure, is the substance behind our
use of the adjective `physical' for these theories. These theories are
still approximate, of course, and they will not have the internal
consistency of statistical mechanical theories obtained by exact
analysis of a mechanical Hamiltonian system.

\subsection*{Quasi-chemical theory}\addcontentsline{toc}{subsection}{Quasi-chemical theory}

The quasi-chemical theory\cite{PrattLR:Quatal,HummerG:Newphe} adapted to
treat hard core solutes\cite{Pratt:2001} gives an explicit structure for
the $\Delta\mu_\mathrm{A}$ formula as in Eq.~\ref{eq:pf}.  That result
can be regarded as a formal theorem
\begin{eqnarray}
\Delta\mu_\mathrm{A}  =RT\ln\left[{1 +  \sum\limits_{m\ge 1} K_m{}
\rho_\mathrm{W}{}^m }\right]~.
\label{hsqca}
\end{eqnarray}
The $K_m{}$ are equilibrium ratios
\begin{eqnarray}
K_n={\rho_\mathrm{{\bar A}W_n} \over \rho_\mathrm{{\bar A}W_{n=0}}
\rho_\mathrm{W}{}^n }~
\label{Kn-ob}
\end{eqnarray}
for binding of solvent molecules to a cavity stencil associated with the
AW excluded volume, understood according to the chemical view
\begin{eqnarray} \mathrm{{\bar A}W_{n=0}\ +\ nW \rightleftharpoons {\bar A}W_n}~.
\label{reaction} \end{eqnarray}
$\mathrm{{\bar A}}$ is a precisely defined cavity
species\cite{Pratt:2001} corresponding to the AW
excluded volume.  Eq.~\ref{eq:pf} should be compared to
Eq.~\ref{hsqca}; because of the structural similarity, it 
is most appropriate to consider Eq.~\ref{eq:pf} as a
quasi-chemical approximation.  The $K_m{}$ are well-defined
theoretically\cite{Pratt:2001} and observable from
simulations. So again this approach can be significantly constructive
when combined with simulations\cite{PrattLR:Quatal,Pratt:ES:99,HummerG:Newphe}.  But
the first utility is that the low density limiting values of $K_m{}$,
call them $K_m{}^{(0)}$, are computable few body
quantities\cite{Pratt:2001}.  The approximation
\begin{eqnarray}
\Delta\mu_\mathrm{A} \approx RT\ln\left[1 +  \sum\limits_{m\ge 1} K_m{}^{(0)}
\rho_\mathrm{W}{}^me^{- m\zeta_1} \right]
\label{pqca}
\end{eqnarray}
is then a simple physical theory, the primitive quasi-chemical
approximation\cite{Pratt:2001}. The Lagrange multiplier $\zeta_1$ serves
as a `mean field' that adjusts the mean occupancy to the thermodynamic
state of interest. For a hard sphere solute A in a hard sphere solvent,
this theory produces sensible results though it does not achieve high
accuracy in the dense fluid regime $\rho d^3> 0.7$\cite{Pratt:2001}
where $d$ is the diameter of the solvent hard spheres. [The foremost
questions for aqueous solutions are at the lower boundary of this
conventional demarcation of dense fluids.]  When the accuracy of this
theory for hard sphere systems degenerates, it is because the
distribution
\begin{eqnarray}
{\hat p}_n={K_{n}{}^{(0)}\rho_\mathrm{W}{}^n e^{- n\zeta_1} 
\over 
 1 +\sum\limits_{m\ge 1} K_m{}^{(0)} \rho_\mathrm{W}{}^me^{- m\zeta_1} }
\label{cluster-var}
\end{eqnarray}
is too broad in the low n extreme\cite{Pratt:2001}; see Fig.~2. Because
this theory thus directly treats short range molecular structure but
somewhat too broadly, it is a natural suggestion for generating default
models.  No direct experience along those lines is presently available.

The investigation of how such a simple theory breaks
Eq.~\ref{cluster-var} down is a yet more interesting aspect of the
development of the quasi-chemical theories for these
problems\cite{HummerG:Newphe,Pratt:2001}.  For the hard sphere fluid at
higher densities the primitive quasi-chemical theory
Eq.~\ref{cluster-var} remains a faithful descriptor of the n$\ge$1
features of the distribution.  But the actual $p_0$ (see Fig.~2) becomes
depressed relative to the model;  $p_0$ breaks away from the rest of the
primitive quasi-chemical distribution. In fact, the suggested
correlation correction can be effectively empiricised and provides an
accurate description of these distributions for hard sphere fluids.
These exotic complexities with $p_0$ are not reflected in
Eq.~\ref{gaussian}.  Tiny features like that were noticed, however, in
the initial simulation studies of these probability
models\cite{Hummer:PNAS:96}. Additionally, we anticipate discussion
below by noting that proximity to a low pressure liquid-vapor transition
point, associated with solvent-solvent attractive interactions and
potential dewetting of hard surfaces, is expected to {\em increase}
$p_0$\cite{Hummer:PNAS:96,HummerG:Newphe}.

\subsection*{Importance sampling to correct occupancy
probabilities}\label{corrections}\addcontentsline{toc}{subsection}{Importance
sampling to correct occupancy probabilities}

A more workman-like investigation of these theories can be based upon
the potential distribution theorem\cite{Widom:JCP:63,Pratt:ECC,Pratt:2001}
\begin{eqnarray}
e^{-\Delta \mu _A/RT}=\left\langle\left\langle {e^{-\Delta U/RT}}
\right\rangle\right\rangle _0~.
\label{widom}
\end{eqnarray}
The brackets $\left\langle\left\langle \ldots \right\rangle\right\rangle
_0$ indicate the average of the thermal motion of a distinguished A solute and the
solvent under the condition of no interactions between these subsystems;
 the latter restriction is conveyed by the subscript `0.'  Distributions
${\hat p}_n$ should be helpful as importance functions.  It would be
natural to use this estimate to revise the calculation of all the
probabilities $p_n$.  But we have seen a case, Fig.~2, where the
distinction between n=0 and n$\ge$1 is most interesting.  In addition,
this quantity averaged here Eq.~\ref{widom} takes the values zero (0) if
n$\ge$1 and one (1) for n=0. Thus we consider the importance function
\begin{eqnarray}
W=\left\{ {\matrix{{\hat p_0,n=0~,}\cr
{1-\hat p_0,n>0.}\cr
}} \right\}
\label{weight}
\end{eqnarray}
The standard importance sampling ideas\cite{Torrie:JCompP:77,Valleau}
then produce
\begin{eqnarray}
e^{-\Delta \mu _A/RT}={{\left\langle \left\langle {W \,e^{-\Delta U/RT}}
\right\rangle\right\rangle _{1/W }} \over {\left\langle \left\langle {W}
\right\rangle\right\rangle _{1/W }}}~.
\label{important}
\end{eqnarray}
The sampling distribution is the Boltzmann weight in Eq.~\ref{widom}
multiplicatively augmented by the configurational function $1/W$ and
corresponds to a finite probability  step $p_0 \over 1 - p_0$ as the
first solvent molecule enters the observation volume.  Typically, this 
will  lead to a diminished occupancy of the observation volume.
After some rearrangement, Eq.~\ref{important} is
\begin{eqnarray}
{ p_0-\hat p_0 \over \hat p_0}=
 {  \left( 2 \pi _0-1 \right) \over  1
  -\pi_0
\left( {1- 2 \hat p_0 \over 1-
\hat p_0 }\right)  }
\label{correct}~.
\end{eqnarray}
Here $\pi_0$ is the probability of the observation volume being empty
with the reweighted sampling
\begin{eqnarray}
\pi_0 = \left\langle \left\langle e^{-\Delta U/RT}
\right\rangle\right\rangle _{1/W }~.
\label{pi0}
\end{eqnarray} These formulae could be used directly with simulation
calculations.  The best available approximate
$p_0$\cite{Stillinger:73,Ashbaugh:2001} could be used as $\hat p_0$ in
order to achieve higher accuracy; that would be interesting but not easy
because achieving n$\gg 0 \rightarrow$ n=0 transitions requires rare
collective processes. On the other hand, n=0$\rightarrow$n$\ge$1
transitions will have a low acceptance probability. But our argument
here is directed toward understanding physical features left out of
simple models such as the quasi-chemical model of Eq.~\ref{eq:pf}.  The
choice here of W in Eq.~\ref{weight} is transparently directed towards
discussion of a `two-state' picture of the hydration.

With that goal, the conceptual perspective is the more interesting one. 
This is a discrete example of a procedure common in density functional
arguments. Local particle occupancies are altered by a reweighting. 
Just as with $p_0$, $\pi_0$ can be studied with a Mayer-Montroll series
and moment modeling\cite{Pratt:NATO99}. The moments involved now would 
be obtained from study of the designed non-uniform system. The average
on the right-side of Eq.~\ref{pi0} is a functional of the density
induced by the reweighting.

For example, if the reweighting can serve to nucleate a bubble because
of proximity of the thermodynamic state to a liquid-vapor transition,
then, physically viewed, $\pi_0$ is expected to be composed of two
important cases:  (a) `vapor' with a depletion region
surrounding the observation volume; this gives contribution one (1) to
$\pi_0$ for these cases and (b)  `liquid;' typically these cases will
contribute zero (0) to $\pi_0$.   But occasional configurations,
roughly with frequency $\hat p_0$,  will give $e^{-\Delta U/RT}=1$.  
Thus as a rough estimate,  we expect
\begin{eqnarray}
\pi_0 \approx {e^{-\Delta F/RT}({1-{\hat p_0} \over {\hat p_0}}) + {\hat
p_0} \over e^{-\Delta F/RT}({1-{\hat p_0} \over {\hat p_0}}) + 1}
\label{engest}
\end{eqnarray}
where $\Delta F$ is the free energy for formation of a bubble from the
liquid corresponding to boundary conditions n=0 on the observation
volume.  That free energy might be approximated by a combination of
van~der~Waals theories and molecular theories appropriate for the vapor
phase\cite{weeks:ARPC2001}.  Let's consider a one phase, dense liquid
thermodynamic state, not far from coexistance with a vapor phase so
that 1$>e^{-\Delta F/RT} > {\hat p_0}$.  With these estimates,
Eqs.~\ref{correct} and \ref{engest} evaluate to
\begin{eqnarray}
p_0 \approx e^{-\Delta F/RT} >{\hat p_0}
\label{bubble}
\end{eqnarray}
The insertion probability is the probability of bubble formation, surely
the only simple guess, and this estimated change raises the value ${\hat
p_0}$. 

\subsection*{van~der~Waals
picture}\addcontentsline{toc}{subsection}{van~der~Waals picture}
On the basis of the observations above, we can construct the following
picture\cite{Pratt:2001} by considering how these theories would work
for an atomic size hard sphere solute in a simple van~der~Waals fluid
system. Consider packing effects first, then subsequently the effects of
attractive interactions. For dense liquid cases with full-blown packing
difficulties, models such as Eq.~\ref{gaussian} or \ref{cluster-var}
overestimate $p_0$ because those models aren't accurate for packing
problems in the dense fluid regime. Next, consider attractive
interactions and the possibility of dewetting. Those effects raise
$p_0$.  Models such as Eq.~\ref{gaussian} do not reflect these
phenomena.  But these two errors can compensate, so Eq.~\ref{gaussian}
can be empirically accurate for atomic solutes despite the naivit\'{e}.
This is another rationalization of the astonishing, fortuitous accuracy
of the two moment model and of the Pratt-Chandler theory.

While absorbing this argument, there are two additional points that may
be noted.  The first point is that in discussing errors in treating
packing effects, we have been concerned about errors of the same type as
those in the Percus-Yevick theory for the hard sphere fluid.  But the
Percus-Yevick theory of the hard sphere fluid might be considered {\em
the} most successful theory of a liquid, ``gloriously accurate,
considering its simplicity''\cite{Stell:77}.  So this discussion is
bringing a high-sensitivity view to this problem.  This is necessary
because of the importance and high interest in these problems. The
second point for note is that `attractive interactions' in this argument
involve solvent-solvent interactions, {\em not} solute-solvent
attractive interactions.  Part of the subtlety of these discussions is
that the approach that offers models such as Eq.~\ref{gaussian} is
sufficiently empirical that a unique identification of the source of a
particular inaccuracy is nontrivial.

\section*{HYDROPHOBIC HYDRATION AND TEMPERATURE
DEPENDENCES}\addcontentsline{toc}{section}{HYDROPHOBIC HYDRATION AND
TEMPERATURE DEPENDENCES}

The most astonishing result of the new theory Eq.~\ref{gaussian} is its
explanation of the hydrophobic temperature dependence known as `entropy
convergence'
\cite{Garde:PRL:96,Pratt:NATO99,Garde:99,HummerG:Newphe,GardeS:Temdhh}
and those temperature dependences are discussed here.

\subsection*{Solubilities}\addcontentsline{toc}{subsection}{Solubilities}

The solubilities of simple gases in water have some interesting
complexities\cite{POLLACKGL:WHYGDL}.  Many simple gases have a
solubility minimum in water at moderate temperatures and pressures. 
Since the solubility is governed by $\Delta\mu_\mathrm{A}/RT$, the
temperature variation of the solubility at a fixed pressure requests
information on
\begin{eqnarray}
\left( {{{\partial \Delta \mu _\mathrm{A}/RT} \over {\partial T}}} \right)_p=-{1
\over T}\left( {{{\Delta h_\mathrm{A}} \over {RT}}} \right)
\end{eqnarray}
with $\Delta h_\mathrm{A}$ the partial molar enthalpy change upon
dissolution of species A and we considering the low concentration limit
here.  Thus, a solubility minimum leads us to anticipate a temperature
of zero enthalpy change for the dissolution where $\Delta \mu
_\mathrm{A}/RT$ plotted as a function of T has a maximum.

At a higher temperature the dissolution of many simple gases shows
approximately zero partial molar entropy change:
\begin{eqnarray}
\left( {{{\partial \Delta \mu _\mathrm{A}} \over {\partial T}}}
\right)_p=- {{{\Delta s_\mathrm{A}}}}~.
\end{eqnarray}
$\Delta \mu _\mathrm{A}$ plotted as a function of $T$ has a maximum.
More puzzling is the fact that this temperature of zero entropy change
is common to a number of different gases.  This phenomenon is referred
to as `entropy convergence' because the entropies of hydration of
different solutes converge to approximately zero at a common
temperature\cite{Baldwin:2001,LEEB:ISOITT}.

\subsection*{Model explanation}\addcontentsline{toc}{subsection}{Model
explanation} The puzzle ``why?'' was first answered on a molecular level
in Ref.~\cite{Garde:PRL:96} on the basis of the model Eq~\ref{gaussian};
see also\cite{Pratt:NATO99,Garde:99,HummerG:Newphe,GardeS:Temdhh}.
Establishing these two temperature behaviors should go a long way toward
establishing the temperature variations of hydrophobic effects
throughout an extended range relevant to biomolecular structure.

If we agree to be guided by an estimate of $p_n$ based upon a continuous
normal distribution\cite{Garde:PRL:96,Pratt:NATO99,Garde:99,%
HummerG:Newphe} then evaluation of the Lagrange multipliers of
Eq.~\ref{gaussian} is not a problem and
\begin{eqnarray}
\Delta\mu_\mathrm{A}/RT &\approx& 
{1\over 2}\left\{ {\langle n \rangle_0{}^2 \over 
\langle \delta n^2 \rangle_0} + 
\ln[2\pi \langle \delta n^2 \rangle_0] \right\} 
\label{eq:muex.a} \\ 
	& = & {1\over 2}\left\{ {\left(\rho v\right) ^2  \over 
\langle \delta n^2 \rangle_0} + 
\ln[2\pi \langle \delta n^2 \rangle_0] \right\}
\label{eq:muex.b} \end{eqnarray}
with $v$ the volume of the AW excluded volume, expected to be weakly
temperature dependent.   The surprise is that $\langle \delta n^2
\rangle_0$ varies only slightly with temperature over the interesting
temperature range.  This is suggestive of the data shown in Fig.~1. 
Furthermore the second term of Eq.~\ref{eq:muex.b} is smaller than the
first.  Therefore the plot of $\Delta \mu _\mathrm{A}$ as a function of
$T$ experiences a maximum because the combination $T \rho^2$ has a
maximum as $\rho=\rho_{sat}(T)$ decreases with increasing temperature
along the coexistence curve.  To the extent that the rightmost term of
Eq.~\ref{eq:muex.b} can be neglected and $v^2/\langle \delta n^2
\rangle_0$ is independent of temperature, then entropy convergence will
occur and the temperature at entropy convergence will be the same for
all hydrophobic solutes.

This development applies to model hard core solutes and the convergence
temperature does appear to shift slowly but systematically to lower
temperatures as the volume of the solute increases.  But it was
disturbingly noted nearly forty years ago that the success of the scaled
particle model in evaluating hydration entropies ``$\ldots$ suggests an
almost thermodynamic independence of molecular
structure''\cite{Ben-Naim:67}.  In the entropy convergence phenomena, we
see that this almost thermodynamic independence of molecular structure
is a feature of the data\cite{LEEB:ISOITT} and that the current theory
gives a simple molecular explanation that resolves that puzzle.

The  lower temperature iso-enthalpy solubility minimum is expected to be
tied to a different aspect of the solution-water interactions, the
van~der~Waals attractive interactions. Following a WCA view, these
effects should be reasonably described by first order perturbation
theory so as a qualitative model we have\cite{Garde:99,HummerG:Newphe}
\begin{eqnarray}
\Delta\mu_\mathrm{A}\approx  - A \rho + B T \rho^2 + C T
\label{schemat}
\end{eqnarray}
with fitting parameters A, B, and C.  This equation does indeed have the
correct qualitative behaviors\cite{Garde:99,HummerG:Newphe}. If this
Eq.~\ref{schemat} is used as a fitting model and the parameters are
unrestricted, it is essentially perfect.   If the parameters are
constrained by physical expectations for the temperature independent
fitted parameters are v, $\langle \delta n^2 \rangle_0$, and A, then
this model is only qualitatively and crudely successful in describing
experimental solubilities.

Several of these considerations have been reexamined recently with
results consistent with this
picture\cite{BoulougourisGC:Hencaw,Ashbaugh:2001b}.  Although these
temperature behaviors were not always so clearly recognized as this,
there remains specific solubility issues that aren't resolved including
at the simulation level.  An interesting case was provided by  the
important simulation calculations of Swope and
Andersen\cite{SWOPEWC:MOLMCS} on solubility of inert gas atoms in water.
With regard to the Lennard-Jones solute water (oxygen) interaction
models, they concluded:  ``For the potentials used in the present
simulations, it is not possible to fit the value, slope, and curvature
for helium and neon without choosing what we believe to be unreasonably
large values of $\sigma$.  We can, however, obtain fits to the data for
argon and krypton with reasonable values of the diameters.'' Simulation
calculations for such cases have been pursued several times since then; 
those activities up to 1998 are summarized by Arthur and
Haymet\cite{ArthurJW:Solnsw} and the latter effort also concludes with
some ambiguity about the case of the He solute.

It is also important to emphasize that this model is used here only over
a limited temperature range and at low pressure.  Lin and
Wood\cite{LinCL:Pretfe} used molecular dynamics to model the
thermodynamic properties of small hydrocarbons in water over a wide
range of temperature and pressure and Errington, {\em et
al.\/}\cite{ErringtonJR:Molspe} studied the phase equilibria of
water-methane and water-ethane systems of over wide ranges of
temperature and pressure using Monte Carlo techniques.

\section*{RECENT EXPERIMENTAL STUDIES OF HYDRATION
STRUCTURE}\addcontentsline{toc}{section}{RECENT EXPERIMENTAL STUDIES OF
HYDRATION STRUCTURE}

In recent years, efforts to measure directly the structure of water
surrounding simple hydrophobic solutes have produced results that should
be of quantitative relevance to the theories discussed
here\cite{DeJongPHK:Hydhm,Filipponi:97,%
BowronDT:Temdth,BowronDT:Hydhfc,BowronDT:Xraasi,SullivanDM:Hydhah}.  A
more quantitative consideration is warranted but the initial impression
is that these results are in good agreement with the calculations that
have been done.  These data give the weight of evidence to important
basic conclusions also.

\subsection*{Pressure Dependence of Hydrophobic
Hydration}\addcontentsline{toc}{subsection}{Pressure dependence of
hydrophobic hydration} One such conclusion is that this structuring of
water appears to be independent of variations of the pressure to 700~bar
(70~MPa)\cite{BowronDT:Xraasi}.  This may be important to the current
issue of pressure denaturation of soluble proteins as is discussed
below. But investigation in a higher pressure range would be necessary
for that purpose\cite{ChauPL:Comsts,GHOSHT:2001a}.

\subsection*{`Clathrate' is in the eye of the
beholder}\addcontentsline{toc}{subsection}{`Clathrate' is in the eye of
the beholder}
Another important conclusion follows from the direct comparison of the
radial distribution of oxygen atoms surrounding Kr in liquid aqueous
solution and in a solid clathrate phase. Those radial distributions are
qualitatively different in the two different phases. This is important
because a  `clathrate' picture of hydration structure of nonpolar
solutes in liquid water is a common  view of hydrophobic hydration that
has not been of quantitative relevance; it has been a `pictorial
theory'\cite{Frank:JCP:45}. In contrast, several theoretical calculations
that have had quantitative value assume roughly the antithesis of
`clathrate,' that the conditional density of water surrounding of a
non-spherical nonpolar solute can be built-up by superposition of
proximal radial information\cite{pellegrinim:modscc,pellegrinim:potmfb,%
GardeS:HydiCe,AshbaughHS:Coneaa,AshbaughHS:EnthhE,Garde:PRE:96}. The
`clathrate' language is widely used, hardly explicitly justified, and
leads to misunderstandings.  A number of studies have explicitly
considered the issue of how valid is the `clathrate'
description\cite{Head-Gordon:PNAS:95,FoisE:Hydecs,%
LAAKSONENA:MOLNMW,TANAKAH:HYDHIT,wallqvista:molsha,ChengYK:Surtdb,MountainRD:Hydsh}.
The conclusion seems to be that if you look for clathrate-style
hydration structures you probably see them but if you ask whether they
are necessary for a correct quantitative understanding, the answer is
`no.'  `Clathrate' is in the eye of the
beholde.  A reasonable recommendation is that when `clathrate' is used as a
descriptor in these liquid solutions, it should be explicitly defined
and justified.  Attempts to formulate quantitative theories on the basis
of chemical models of these hydration shells are known but
ill-developed\cite{HummerG:Newphe}.

\section*{POTENTIALS OF THE MEAN FORCES AMONG PRIMITIVE HYDROPHOBIC
SPECIES IN WATER}\label{pmfs}\addcontentsline{toc}{section}{POTENTIALS
OF THE MEAN FORCES AMONG PRIMITIVE HYDROPHOBIC
SPECIES IN WATER}

The theories discussed above are straight-forwardly applicable to
non-spherical solutes. For a solute composed of two atoms with varying
interatomic separation, the comparison
\begin{eqnarray}
\Delta\mu_{AA'}(r) - \Delta\mu_A - \Delta\mu_{A'} \equiv w_{AA'}(r)
\label{pmf}
\end{eqnarray}
leads to the classic issue of the `potential of mean
force' (pmf), an issue of long-standing interest\cite{DASHEVSKYVG:SOLHIN,PANGALICS:DETTM2,%
Pangali:79,swaminathans:monchb,ravishankerg:moncst,BEVERIDGEDL:FRESAT,%
RAVISHANKERG:POTMFS,backxp:signrf,backxp:somcrr,WATANABEK:MOLSTH,%
watanabe:86,WALLQVISTA:HYDIBM,JORGENSENWL:EFFCAF,BERNEBJ:MODHI,%
JORGENSENWL:AAPMF,LINSEP:ORIBBP,LINSEP:STATBD,LINSEP:MOLSDA,%
Smith:JACS:92,SKIPPERNT:COMMWS,VANBELLED:MOLSMH,Smith:JCP:93,%
danglx:potmfm,newmh:molcte,LudemannS:inftph,Ludemann:97,Payne:97,%
youngws:reethe,Rick:97,RankJA:Conshb,ChipotC:Bendgm,Rick:2000,%
ShimizuS:Temdhi,Gervazi:AAPMF,GHOSHT:2001a,GHOSHT:2001b}. The model Eq.~\ref{gaussian}
for a simple case was tested against simulation
results\cite{Hummer:PNAS:96} and the comparison was close.
That theory, including Eq.~\ref{eq:muex.b}, sheds new light on these
properties.

\subsection*{Contact Hydrophobic
Interactions}\addcontentsline{toc}{subsection}{Contact hydrophobic
interactions}
Consider $w_{AA}(r)$ for atomic solutes in contact. When the solute
atoms are in van~der~Waals contact, we still anticipate that $\langle
\delta n^2 \rangle_0$ will be only weakly temperature dependent. 
Further, the volume excluded to the solvent by the pair is less than
twice the volume of an atom alone because of the overlap of their
excluded regions.   Thus the dominant contribution to Eq.~\ref{pmf} is
negative, stabilizing the contact pair, and that stabilization increases
with temperature below the entropy convergence temperature.

[Note that the subtraction Eq.~\ref{pmf} requires some additional
thought when the logarithmic term of Eq.~\ref{eq:muex.b}   is
addressed\cite{Hummer:JPCB:98}.  If the subtraction were naively
carried-out, that r-independent nonzero difference might imply
pathologically long-ranged interactions.  A more careful consideration
of the statistical approaches satisfactorily resolves that
pathology\cite{Hummer:JPCB:98}.  This detail shows again that these
theories are not naively equivalent to the Pratt-Chandler theory.]

\subsection*{Non-contact Hydrophobic
Interactions}\addcontentsline{toc}{subsection}{Non-contact hydrophobic
interactions} On the other hand, when the atomic solutes are separated
enough that a water molecule may fit between them, the volume excluded
to the solvent by the pair is more nearly twice the excluded volume of
the separated atoms. The theories following Eq.~\ref{gaussian} then are
more sensitive to water molecule correlations of longer range because
the information $\langle \delta n^2 \rangle_0$ depends on those
correlations.  These theories then produce more subtle effects. 
Non-contact hydrophobic interactions may be, nevertheless, significant
because of the larger configurational volume corresponding to those
solvent-separated configurations.  In addition, as discussed by Pratt
and Chandler\cite{Pratt:80b,Pratt:85a}, free energies of these
solvent-separated configurations may be more sensitive to details of
van~der~Waals attractive interactions than are contact configurations.

\subsection*{Simulation
results}\addcontentsline{toc}{subsection}{Simulation results}

These views seem to be born out by the available simulation results,
although\cite{LudemannS:inftph} ``the conclusions drawn from previous
simulation calculations have been very contradictory.''  Substantial
stability for contact hydrophobic pairs is probably the least
contradictory of the possible conclusions.  The recent preponderance of
simulation results indicate that these contact pairs are stabilized by
favorable hydration entropies\cite{Smith:JACS:92,Smith:JCP:93,%
SKIPPERNT:COMMWS,danglx:potmfm,LudemannS:inftph,Ludemann:97,Rick:97,%
HummerG:FastiE,GHOSHT:2001b}. This would agree with the view established
from the simple model Eq.~\ref{eq:muex.b} but those model temperature
variations have been checked mostly along the liquid-vapor coexistence
curve.  The interesting results of
Ref.~\cite{LudemannS:inftph,Ludemann:97} find substantial temperature
variations at fixed water density.  It was noted\cite{danglx:potmfm}
that the conditions of temperature increase at fixed density strengthen
the apparent entropic stabilization of the contact pair.  That is also
how Eq.~\ref{eq:muex.b} works; the density decrease serves to moderate
the temperature increase and eventually, at the entropy convergence
temperature, to dominate it.  But the large effects seen by the Vienna
group\cite{LudemannS:inftph,Ludemann:97} make it unclear that the simple
model Eq.~\ref{eq:muex.b} will be accurate for those phenomena.

Simulation results for the solvent-separated configurations are less
clear also.  There are entropic and enthalpic temperature effects in
opposite directions with small net
results\cite{Smith:JCP:93,Ludemann:97,Rick:97}.   Because of the larger
configurational volume for the solute pair in this configuration, these
smaller hydration free energies are not negligible and\cite{Ludemann:97}
``the puzzling finding that the marked hydrophobic behavior of
methane-like solutes concluded from the free energy data is not
reflected in a similarly clear manner by the second osmotic virial
coefficients requires a closer inspection of the underlying phenomena.''
See also \cite{Rossky:80}.  There is precedent, depending on a variety
of additional details, for simulation results to exhibit either
hydrophobic clustering or not, {\it i.e.\/}, ``hydrophobic
repulsion.''\cite{watanabe:86}.

\subsection*{Polarizability?}\addcontentsline{toc}{subsection}{Polarizability?}

The complications suggested for these solvent-separated configurations
seem to have lead to other contradictory results.   It was suggested
long ago\cite{backxp:signrf} that solute polarizability might change the
character of hydrophobic interactions predicted by simple theories.   A
later simulation calculation that included explicit
polarizability\cite{VANBELLED:MOLSMH} in the water-water interactions
also suggested that these more complicated descriptions might
qualitatively change the hydration of non-contact atom pairs. Further
calculations again suggested that polarizability could lead to
substantial changes but in a different direction from those seen
earlier\cite{newmh:molcte}; treatment of long-ranged interactions was
noted as a significant issue in these calculations. More recent studies
of interaction models that include polarizability, however, have
restored an original `small change' view for the moment\cite{Rick:97}.

\subsection*{Alkane conformational equilibrium in
water}\addcontentsline{toc}{subsection}{Alkane conformational
equilibrium in water}

The potentials of mean force just discussed are relevant to
consideration of the conformational equilibrium of small flexible
hydrophobic molecules in water.  The first test case for theories has
always been the {\em trans-gauche\/} isomerization of
n-butane\cite{jorgensen_1982,berne7_1982,Buckner:87,%
Tobias:90,pellegrinim:modscc,Wallqvist:95,WallqvistA:theoth,%
Garde:96:b,AshbaughHS:EnthhE,AshbaughHS:Hydces,AshbaughHS:Coneaa}.  For
the case of n-butane,  solvent separated possibilities are not
available, so the contact hydrophobic interactions are relevant.  The
population of the more compact {\em gauche\/} configuration is enhanced
by an entropic hydration effect.  Hummer, {\it et al.,\/}\cite{Hummer:PNAS:96}
applied to the model of Eq.~\ref{gaussian} to the case of conformational
equilibrium of n-butane in water and found close agreement with the
latest simulation results.  Hummer has effectively adapted the model of
Eq.~\ref{gaussian} so that it can be simply applied to other
alkanes\cite{HummerG:Hydffa}.    Much longer chain molecules that are
strictly hydrophobic as less well studied primarily because they would
be so unusual as isolated components of aqueous solutions.   Hydrophilic
groups are necessary to solubilize large molecules.  Perhaps the
simplest such soluble  molecules would be polyethylene oxide chains
which are a specific interest\cite{BorodinO:moldss}.  But Gallicchio,
{\it et al.,\/} \cite{GallicchioE:Entcda} has recently studied the
hydration of slightly larger alkanes in additional detail.

\section*{PRESSURE DEPENDENCE OF HYDROPHOBIC
INTERACTIONS}\addcontentsline{toc}{section}{PRESSURE DEPENDENCE OF
HYDROPHOBIC INTERACTIONS}

The intense current interest in pressure studies of protein
structure is due to the alternative light that this research can shed on
protein conformational dynamics.  A recent example can be found
in \cite{AkasakaK:Lowesp} but we are unable to  review adequately that
body of interesting work here.   

The complications of the non-contact hydrophobic interactions mentioned
above appear to be involved in understanding pressure denaturation of
proteins, however.  We identify some of that work because views of
hydrophobic effects had been paradoxical for these
issues\cite{{KAUZMANNW:PROSTU}} and because it gives additional
perspective into the theories discussed here.  Wallqvist reported
initial studies of pressure dependence of hydrophobic
interactions\cite{wallqvista:molsha,wallqvista:premsa}. Remarkably,
atomic hydrophobic solutes in water clustered at low pressure but
dispersed at a substantially higher pressure that wasn't further
quantified in that study.  Later the Rutgers group took an important
step\cite{Payne:97} in Monte Carlo calculations of the effects of
pressure on the pmf between  Lennard-Jones model hydrophobic atomic
solutes in water.  Hummer, {\it et al.\/}\cite{Hummer:PNAS:98} then
developed the theory Eq.~\ref{gaussian} for these pmfs as a function of
pressure. That theory suggested that non-contact configurations of
hydrophobic pairs become progressively more stable relative to contact
pairs and that this might be a feature of pressure denaturation that was
known to produce less disrupted structures than does heat denaturation.
Subsequent molecular dynamics calculations\cite{GHOSHT:2001a} confirmed
this picture of dispersal at higher pressures and the pressure
variations of the pmfs: as pressure is increased, these pmfs become more
structured, the contact minimum deepens, the desolvation barrier becomes
higher, the solvent-separated well becomes better defined, deeper,   and
it appears to deepen faster than the contact well.  This work also
observed clustering of hydrophobic atomic solutes at low pressure
(1~atm) but dispersal at high pressure (8000~atm, 810.6~MPa). This work
also saw changes in the solute-water(oxygen) radial distribution that
should be observable in experimental studies such as those of
Ref.~\cite{BowronDT:Xraasi}.  But the simulation results are for
considerably higher pressures than the experimental work reported. These
calculations also considered spherical hydrophobic solutes of a larger
size,  more comparable with valine, leucine, or iso-leucine side chains.
The responses to substantial pressure increases were similar but perhaps
slightly more pronounced. A physical view is that as the pressure is
increased, water molecules can be jammed between contact hydrophobic
pairs; this evidently results in a more efficient, lower-volume packing,
and consequently a negative hydration free energy change with increasing
pressure for non-contact configurations.

\section*{SIZE DEPENDENCE OF HYDROPHOBIC HYDRATION FOR HARD SPHERE
SOLUTES}\addcontentsline{toc}{section}{SIZE DEPENDENCE}
For a hard sphere solute, the rate of increase of the hydration free
energy Eq~\ref{p0} with the distance of closest AO approach, denoted by
$\lambda$, produces a particularly interesting quantity 
\begin{eqnarray}
\rho_W G(\lambda )={1 \over {4\pi \lambda ^2 }}\left( {{{\partial \Delta
\mu _A/RT} \over {\partial \lambda }}} \right)~. 
\end{eqnarray} 
This $\rho_W G(\lambda)$ is the conditional density of the solvent water
(oxygen) at contact with the spherical solute.  Because of the
involvement of $\partial\Delta\mu_A/\partial \lambda$ this relation
describes the compressive force exerted by the solvent on the solute.

\subsection*{Contact densities}\addcontentsline{toc}{subsection}{Contact
densities} Direct studies of these quantities have
shown\cite{Pratt:PNAS:92,Palma} that in the range
2.0~\AA$<\lambda<$3.0~\AA\ $G(\lambda)$ for liquid water is
approximately two-times larger than for n-hexane. Water exerts a higher
compressive force on the surface of an inert solute than do typical
organic liquids so that water squeezes-out hydrophobic solutes
\cite{Richards:SA:91}.  More pertinent for the present discussion is
that the Pratt-Chandler theory overestimates this compressive force and
the original scaled particle model underestimates
it\cite{Pratt:PNAS:92,Palma}.  The revised scaled particle
model\cite{Stillinger:73} lands in the middle and does a better job
at describing this compressive force.  Recent work\cite{Ashbaugh:2001}
has studied these quantities over a much larger range of $\lambda$ and
confirmed the accuracy of the revised scaled particle model.

The reason for the differences between the scaled particle models and
the Pratt-Chandler theory is associated with the known behavior $\rho_W
G(\lambda) \sim p/RT$ for large $\lambda$.  For the cases of first
interest, $p/\rho_W RT\ll$1. Since  $G(\lambda)$ is initially one (1) and
typically increases initially, $G(\lambda)$ decreases for large
$\lambda$ to achieve the small value $p/\rho_WRT$.  In fact, this
decreasing behavior obtains for $\lambda >$3~\AA, approximately.  Thus,
for hard sphere solutes with $\lambda\gg$3~\AA\ the contact density can
be small. This low pressure for a dense liquid is due to attractive
forces between the solvent molecules.  If the conditions are adjusted
for liquid-vapor coexistence, then p is also the pressure of the
coexisting vapor and $p/RT$ would be the density of the vapor under the
assumption that it can be treated as ideal.  Under these conditions we
can, therefore, say that a sufficiently large hard sphere solute
nucleates a bubble of the vapor.  These behaviors are built into the
approximate scaled particle models but not into the Pratt-Chandler
theory. These issues had been directly investigated for spherical model
solutes moderately larger than canonical for
methane\cite{Hummer:PRL:98}; weak effects and were found for those cases
and convincing models were developed\cite{Hummer:PRL:98}.

\subsection*{PMF for stacked plates in
water}\addcontentsline{toc}{subsection}{PMF for stacked plates in water}
This issue has been of particular interest recently because a previous
calculation\cite{Wallqvist:JPC:95:b} studied the pmf between modeled
stacked plates with exclusively repulsive interactions with water
molecules.  That work suggested that contact hydrophobic interactions in
that case could be dominated by a dewetting event: the last two layers
of water molecules intervening between  parallel plates evacuated together. 

\subsection*{Benzene-Benzene
PMF}\addcontentsline{toc}{subsection}{Benzene-benzene PMF}

The comparable results for more realistically modeled benzene, or
toluene, or other small aromatic solute molecule pairs are also
interesting but more
complicated\cite{Rossky:80,JORGENSENWL:AAPMF,LINSEP:ORIBBP,%
LINSEP:MOLSDA,LINSEP:STATBD,GAOJL:SUPHOP,ChipotC:Bendgm,Gervazi:AAPMF}.
These molecules are slightly smaller than the stacked plates that were
studied. In the first place, `T'-shaped contact pairs are more probable
for benzene than a stacked arrangement is. The opposite is true for
toluene\cite{ChipotC:Bendgm}. This is also true for the gas-phase
potential energy surface though the hydration seems to enhance this
distinction\cite{ChipotC:Bendgm} slightly. In the second place,
variations in these
pmfs\cite{Rossky:80,LINSEP:ORIBBP,LINSEP:MOLSDA,LINSEP:STATBD,%
ChipotC:Bendgm,Gervazi:AAPMF} are much smaller  than for the modeled
stacked plates\cite{Wallqvist:JPC:95:b}. A dewetting transition is not
obvious for the calculations with higher molecular realism.  It may be
the significance of any dewetting would be more obvious near
transitional configurations such as the desolvation barrier region that
separates contact from solvent-separated configurations.  Though the
hydration of neither of the high probability configurations discussed
here seemed\cite{LINSEP:ORIBBP,LINSEP:MOLSDA,%
LINSEP:STATBD,ChipotC:Bendgm,Gervazi:AAPMF} remarkable in this way, the
variation of the free energy in the de\-solvation barrier region might
be unusual;  this deserves further checking.  How these complications
are affected by the more complicated environment of an amino acid side
chain, {\it e.g.,\/} phenyl alanine, in a hydrated protein is not known;
the peptide backbone is, of course, highly polar. The hydrogen bonding
possibilities of tyrosine or tryptophan side chains complicate things
yet again. An interesting study of pairing of tryptophan-histidene side
chains \cite{Gervazi:AAPMF} suggested that hydration of these side
chains results in stacked pairing near protein surfaces but `T'-contacts
in protein interiors.  [Continuum dielectric models did not provide a
rationalization of that observed tendency\cite{Gervazi:AAPMF}.]  The
variety of the results obtained suggested\cite{Gervazi:AAPMF} ``$\ldots$
the importance of the atomic details of the solvent in determining the
free energy for the solute-solute interactions.''

\subsection*{Theory of interface
formation\cite{weeks:ARPC2001}}\addcontentsline{toc}{subsection}{Theory of interface
formation}
The development of the theory corresponding to the stacked plates data
was initiated by  Lum, Chandler, and Weeks; see Fig.~1
there\cite{Lum:JPCB:99}. Subsequently\cite{HuangDM:Scahsf,%
HuangDM:Temlsd,HuangDM:Cavfdt,SunSX:Weidft}, a focus has been the study
of how the entropy dominated hydration free energies discussed above for
inert gas solubilities change to the surface tension dominated behavior
expected for ideal mesoscopic hydrophobic species; see
also\cite{SouthallNT:mechsd}.  That behavior was explicitly built into
the revised scaled particle model for hard sphere solutes decades
ago\cite{Stillinger:73}; that model is known to be accurate for
water\cite{Ashbaugh:2001} and for a simple
liquid\cite{HuangDM:Scahsf,HuangDM:Cavfdt}. A fundamental niche for that
theoretical work\cite{weeks:ARPC2001} is the development of a molecular
description of the interface formation mechanism built into the revised
scaled particle model at large sizes\cite{weeks:ARPC2001}.

How these theoretical developments will accommodate heterogeneity of
chemistry and structure that is typical of biomaterials, in contrast to
the model stacked plates, is not yet established.  A nice example of the
issue of heterogeneity, absorption of water on activated carbon, was
discussed recently by M\"{u}ller and Gubbins\cite{MullerEA:Molssh}.  It
is unreasonable to imagine that the biophysical applications will be
simpler than this.  It might be more appropriate at this stage of
development to regard that Berkeley project as ambitiously directed
toward an implicit hydration model\cite{Pratt:ES:99} rather than an
assertion of specific physical relevance of `drying' to biomolecular
structure\cite{ReintenWolde:2001}. These problems require consideration
of several distinct issues together.  One such issue is the direct
contributions of solute-solvent attractive interactions to hydration
free energies for a specific structure\cite{Huang:2001}.   A second
issue is the indirect effects of solute-solvent interactions in
establishing structures and switching between structures as was
initially anticipated\cite{Lum:JPCB:99}.  That switching can be
sensitive to details of van~der~Waals attractive
interactions\cite{BrovchenkoI:Gibesw,WallqvistA:modsdh,BrovchenkoI:Phaewc,%
Ashbaugh:2001,Hummer:2001}.

It is worthwhile attempting to articulate a down-to-earth view  of the
claims of Ref.~\cite{Lum:JPCB:99} specifically. For biomolecules, there
likely are uncommon transitional structures and conditions for which
localized water occupancies can change abruptly.  As these transitional
structures become indentified, they will be interesting. Considering
water-hydrocarbon liquid interfaces, not compromised by hydrophilic
contacts, it is likely that these interfacial regions will be looser
than adjoining bulk phases and more accomodating to imposition of
hydrophobic species\cite{POHORILLEA:MOLAI}.   Surfactants probably
change that conclusion qualitatively\cite{POHORILLEA:MOLAI}.  This is
likely to be relevant to protein hydration and function. Most solute
configurations, except for a few transitional structures, won't require
specific acknowledgement of `drying.' The claims to
Ref.~\cite{Lum:JPCB:99} don't seem to require modification of the
discussion above on POTENTIALS OF THE MEAN FORCES AMONG PRIMITIVE
HYDROPHOBIC SPECIES IN WATER that separated contact from non-contact
configurations and entropy effects from the rest, and then suggested
that the more poorly understood non-contact questions are likely to show
the most variability. The specific claims of Ref.~\cite{Lum:JPCB:99} and
the general issues remain questions for research.

\section*{CONCLUDING
DISCUSSION}\addcontentsline{toc}{section}{CONCLUDING DISCUSSION} This
review has adopted a narrow theoretical focus and a direct style with
the goal of identifying primitive conclusions that might assist in the
next stage of molecular research on these problems.  These theory and
modeling topics haven't been reviewed with this goal recently and a
review of the bigger topic of hydrophobic effects would not be feasible
in this setting. More comprehensive and formal reviews of these topics
are in progress and that must be my excuse for considering such a small
subset of the work in this area.  Nevertheless, some historical
perspective is necessary in identifying valuable primitive conclusions.

One such conclusion is the rectification of the antique
`Pratt-Chandler theory'\cite{Lee:85,Pohorille:JACS:90,%
Pratt:PNAS:92,Palma,Chandler:PRE:93,Hummer:PNAS:96,Garde:PRL:96,%
Pratt:ECC,Hummer:PNAS:98,%
Hummer:PRL:98,Hummer:JPCB:98,Pohorille:PJC:98,Gomez:99,Garde:99,%
Pratt:NATO99,HummerG:Newphe}.  This was an unexpected development
because the advances reviewed above had bypassed stock integral equation
theories. It may have been the stock aspect of the earlier approach
\cite{Pratt:JCP:77} that caused the greatest confusion on this topic.
There is no obvious point to doing that type of integral equation
approximation again for the more complex solutes to which the
theoretical interest has progressed. For the problems addressed,
however, this amended Pratt-Chandler theory is now seen to be a
compelling, approximate theory with empirical ingredients.  The research
noted above on {\it Theory of interface formation\/} and  {\it
Importance sampling to correct quasi-chemical models\/} emphasize that
the treatment of those attractive force effects on the hydration problem
by the Pratt-Chandler theory  was less satisfactory than that of the
scaled particle models.

Another primitive conclusion is that the scaled particle
models\cite{Pierotti:76,Stillinger:73} have been the most valuable
theories for primitive hydrophobic effects.   This is due to the
quantitative focus of those models.  The quantitative focus of the
scaled particle models permitted more incisive
analyses\cite{Ben-Naim:67,Stillinger:73,LUCASM:SIZETN,Lee:85,Pohorille:JACS:90,%
Pratt:PNAS:92,Palma}, in contrast to `pictorial theories,' and those
analyses have lead to significant advances in understanding of these
problems.  The connection from scaled particle models, to Mayer-Montroll
series\cite{Stell:85,Pratt:NATO99},
the potential distribution theorem\cite{Widom:JCP:63,Pratt:ECC}, and the
quasi-chemical approach\cite{PrattLR:Quatal,Pratt:ES:99,%
Rempe:JACS:2000,Rempe:FPE:2001,HummerG:Newphe,Pratt:2001} identifies a
promising line for further molecular theoretical progress on these
problems.  Comparing Eqs.~\ref{eq:pf} and \ref{hsqca},  the
amended Pratt-Chandler theory is most appropriately viewed as a
quasi-chemical theory.  The anticipated theoretical progress will treat
more thoroughly the effects of changes in temperature, pressure, and
composition of the solution, including salt effects, and will treat
neglected `context' hydrophobicity\cite{Rossky:80} in detail. That work
will study cold denaturation and chemical denaturation in molecular
detail.  That work will begin to discriminate hydrophobic effects in the
cores of soluble proteins from hydrophobic effects in membranes and
micelles. That work will begin to consider hydrophobic effects in
nanotechnology with molecular specificity.

A natural explanation of thermodynamic signatures of hydrophobic
hydration, particularly entropy convergence, emerges from these
theoretical advances.   How those temperature behaviors are involved in
cold denaturation or the stability of thermophilic proteins will be a
topic for future research.

Much has been made of the `Gaussian' character of results such as
Fig.~2. The  observation\cite{Hummer:PNAS:96}  best supporting this view
can be explained as a cancellation of approximation
errors\cite{Pratt:2001}; slight inaccuracies of a Percus-Yevick
(`Gaussian') approximation are balanced by neglect of incipient
interface formation for atomic sized solutes.  In a number of other cases
where these distributions have been investigated carefully simple
parabolic models of results such as Fig.~2 are less accurate for
thermodynamic properties and not only because of the influence of a
second thermodynamic phase nearby.  Nevertheless, quadratic models
provide convenient, reasonable starting points for these analyses.

A final conclusion regards the better discrimination of
contact and non-contact hydrophobic interactions.  The contact
hydrophobic interactions seem to express the classic picture of entropy
dominance at lower temperatures. The non-contact interactions have more
variability and are likely to be involved in more unusual effects such
as pressure denaturation where a historical picture of hydrophobicity
had been paradoxical.

\section*{ACKNOWLEDGMENTS}\addcontentsline{toc}{section}{ACKNOWLEDGMENTS}

The `theoretical collaboration at Los Alamos' mentioned in the
INTRODUCTION has included G. Hummer (NIH), A.~E. Garc\'{i}a (LANL), S.
Garde (Rensselaer Polytechnic Institute), M.~A. Gomez (Vassar College),
R.~A. LaViolette (INEEL), M.~E. Paulaitis (Johns Hopkins University),
and A. Pohorille (NASA Ames Research Center). I thank those
collaborators for their numerous essential contributions that I have
discussed in a personal way here. I thank H.~S. Ashbaugh and M.~E.
Paulaitis for helpful discussions of this review. This work was
supported by the US Department of Energy under contract W-7405-ENG-36
and the LDRD program at Los Alamos. LA-UR-01-4900.

\vfill
\pagebreak

\section*{FIGURE LEGENDS}

\begin{description}
\item[Figure 1] The isothermal compressibilities $\beta _T\equiv -\left( {{1
\over \rho }} \right)\left( {{{\partial \rho } \over {\partial p}}}
\right)_T$ along the liquid-vapor coexistence curve of several organic
solvents compared to water\cite{Rowlinson:Swinton:82} in common units. 
From top to bottom are n-heptane, carbon tetrachloride, benzene, and
water.  The compressibility is smaller for water  than for these organic
solvents, is less strongly temperature dependent, and has a minimum
near 46$^\circ$~C.  The critical temperatures of these organic
solvents are all substantially less than the critical temperature
of liquid water.

\item[Figure 2] Distribution ${\hat p}_n$ for a isodiameter hard sphere solute
in a hard sphere fluid at density $\rho d^3$=0.8 and various models. The
dots are simulation results\cite{Pratt:2001} and the spread indicates
a 67\% confidence interval.  The upper dashed curve is the Poisson
distribution with the required mean $\langle n \rangle_0 = 4 \pi \rho
d^3/3$.  The long-dashed curve next down is the primitive
quasi-chemical model, Eq.~\ref{cluster-var}\cite{Pratt:2001}. The solid
line is an `iterated' quasi-chemical theory that incorporates an
empirical correlation correction\cite{Pratt:2001} and provides a simple
accurate description of these quantities for the hard sphere fluid. Note
that the primitive quasi-chemical approximation is a good description
of this distribution for n$\ge$1 but significantly overestimates
$p_0$ at these densities.

\end{description}

\vfill
\pagebreak

\begin{figure}
\begin{center}
\leavevmode
\includegraphics[scale=1.5]{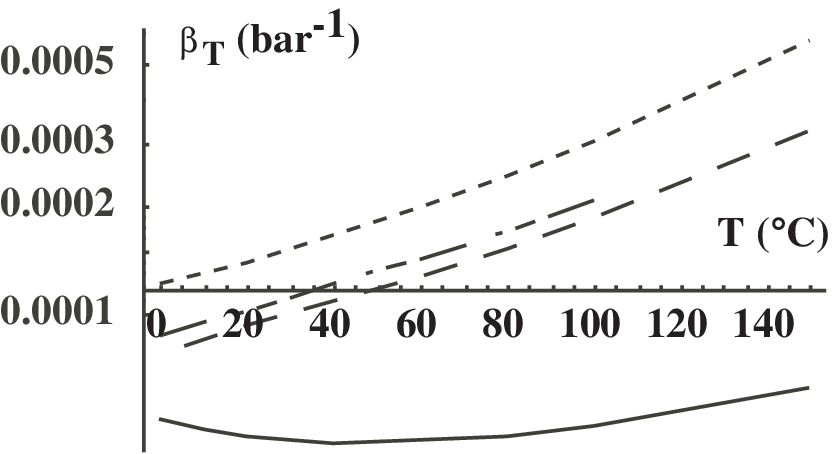}
\end{center}
\caption{}
\label{fig1}
\end{figure}

\vfill
\pagebreak

\begin{figure}
\begin{center}
\leavevmode
\includegraphics[scale=1.2]{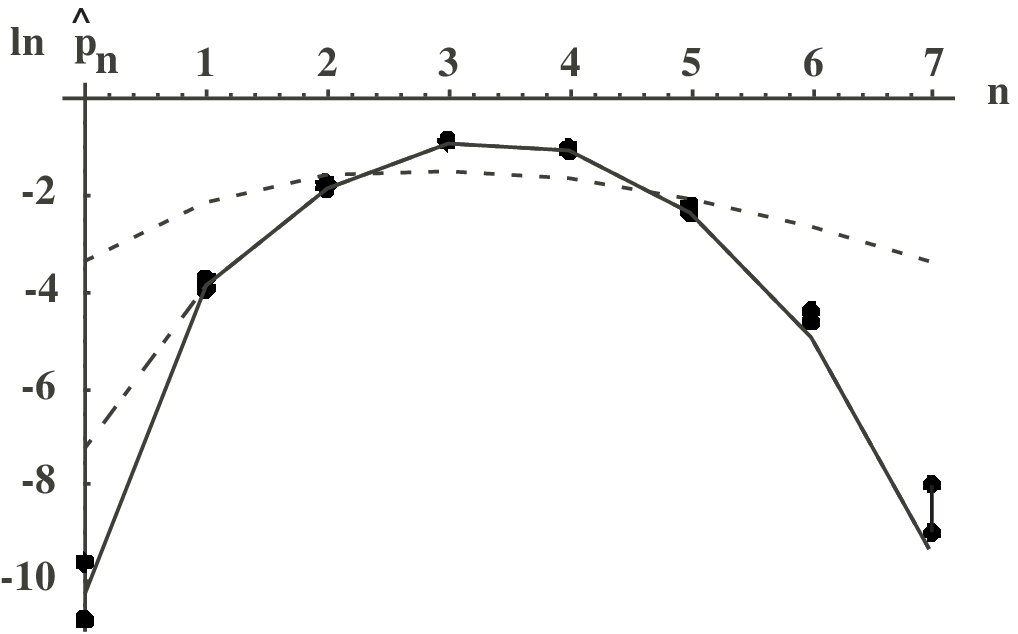}
\end{center}
\caption{}
\label{fig2}
\end{figure}


\begin{thebibliography}{99}

\expandafter\ifx\csname bibnamefont\endcsname\relax
  \def\bibnamefont#1{#1}\fi
\expandafter\ifx\csname bibfnamefont\endcsname\relax
  \def\bibfnamefont#1{#1}\fi
\expandafter\ifx\csname url\endcsname\relax
  \def\url#1{\texttt{#1}}\fi
\expandafter\ifx\csname urlprefix\endcsname\relax\def\urlprefix{URL }\fi
\providecommand{\bibinfo}[2]{#2}
\providecommand{\eprint}[2][]{\url{#2}}

\bibitem{Hummer:PNAS:96}
\bibinfo{author}{\bibnamefont{Hummer} \bibfnamefont{G}},
  \bibinfo{author}{\bibnamefont{Garde} \bibfnamefont{S}},
  \bibinfo{author}{\bibnamefont{Garc{\'{\i}}a} \bibfnamefont{AE}},
  \bibinfo{author}{\bibnamefont{Pohorille} \bibfnamefont{A}},
  \bibinfo{author}{\bibnamefont{Pratt} \bibfnamefont{LR.}} \bibinfo{year}{1996}.
  \bibinfo{journal}{{\it Proc. Natl. Acad. Sci. USA\/}} \bibinfo{volume}{93}:\bibinfo{pages}{8951}

\bibitem{Garde:PRL:96}
\bibinfo{author}{\bibnamefont{Garde} \bibfnamefont{S}},
  \bibinfo{author}{\bibnamefont{Hummer} \bibfnamefont{G}},
  \bibinfo{author}{\bibnamefont{Garc{\'{\i}}a} \bibfnamefont{AE}},
  \bibinfo{author}{\bibnamefont{Paulaitis} \bibfnamefont{ME}},
  \bibinfo{author}{\bibnamefont{Pratt} \bibfnamefont{LR.}} \bibinfo{year}{1996}.
  \bibinfo{journal}{{\it Phys. Rev. Lett.\/}} \bibinfo{volume}{77}:\bibinfo{pages}{4966}

\bibitem{Pratt:ECC}
\bibinfo{author}{\bibnamefont{Pratt} \bibfnamefont{LR.}} \bibinfo{year}{1998}. 
  \emph{\bibinfo{title}{Encyclopedia of Computational Chemistry:``Hydrophobic effects''}} \bibinfo{publisher}{John Wiley
  \& Sons}, \bibinfo{address}{Chichester}
  \bibinfo{pages}{1286--1294}.

\bibitem{Hummer:PNAS:98}
\bibinfo{author}{\bibnamefont{Hummer} \bibfnamefont{G}},
  \bibinfo{author}{\bibnamefont{Garde} \bibfnamefont{S}},
  \bibinfo{author}{\bibnamefont{{Garc\'{\i}a}} \bibfnamefont{AE}},
  \bibinfo{author}{\bibnamefont{Paulaitis} \bibfnamefont{ME}},
  \bibinfo{author}{\bibnamefont{Pratt} \bibfnamefont{LR.}} \bibinfo{year}{1998}.
  \bibinfo{journal}{{\it Proc. Natl. Acad. Sci. USA\/}} \bibinfo{volume}{95}:\bibinfo{pages}{1552}

\bibitem{Hummer:PRL:98}
\bibinfo{author}{\bibnamefont{Hummer} \bibfnamefont{G,}}
  \bibinfo{author}{\bibnamefont{Garde} \bibfnamefont{S.}} \bibinfo{year}{1998}.
  \bibinfo{journal}{{\it Phys. Rev. Lett.\/}} \bibinfo{volume}{80}:\bibinfo{pages}{4193}

\bibitem{Hummer:JPCB:98}
\bibinfo{author}{\bibnamefont{Hummer} \bibfnamefont{G}},
  \bibinfo{author}{\bibnamefont{Garde} \bibfnamefont{S}},
  \bibinfo{author}{\bibnamefont{Garc\'{i}a} \bibfnamefont{AE}},
  \bibinfo{author}{\bibnamefont{Paulaitis} \bibfnamefont{ME}},
  \bibinfo{author}{\bibnamefont{Pratt} \bibfnamefont{LR.}} \bibinfo{year}{1998}.
  \bibinfo{journal}{{\it J. Phys. Chem. B\/}} \bibinfo{volume}{102}:\bibinfo{pages}{10469}

\bibitem{Pohorille:PJC:98}
\bibinfo{author}{\bibnamefont{Pohorille} \bibfnamefont{A.}} \bibinfo{year}{1998}.
  \bibinfo{journal}{{\it Pol. J. Chem.\/}} \bibinfo{volume}{72}:\bibinfo{pages}{1680}

\bibitem{Gomez:99}
\bibinfo{author}{\bibnamefont{Gomez} \bibfnamefont{MA}},
  \bibinfo{author}{\bibnamefont{Pratt} \bibfnamefont{LR}},
  \bibinfo{author}{\bibnamefont{Hummer} \bibfnamefont{G}},
  \bibinfo{author}{\bibnamefont{Garde} \bibfnamefont{S.}} \bibinfo{year}{1999}. \bibinfo{journal}{{\it J.
  Phys. Chem. B\/}} \bibinfo{volume}{103}: \bibinfo{pages}{3520}
 

\bibitem{Garde:99}
\bibinfo{author}{\bibnamefont{Garde} \bibfnamefont{S}},
  \bibinfo{author}{\bibnamefont{Garc\'{i}a} \bibfnamefont{AE}},
  \bibinfo{author}{\bibnamefont{Pratt} \bibfnamefont{LR}},
  \bibinfo{author}{\bibnamefont{Hummer} \bibfnamefont{G.}} \bibinfo{year}{1999}.
  \bibinfo{journal}{{\it Biophys. Chem.\/}} \bibinfo{volume}{78}:\bibinfo{pages}{21}

\bibitem{Pratt:NATO99}
\bibinfo{author}{\bibnamefont{Pratt} \bibfnamefont{LR}},
  \bibinfo{author}{\bibnamefont{Hummer} \bibfnamefont{G}},
  \bibinfo{author}{\bibnamefont{Garde} \bibfnamefont{S.}} \bibinfo{year}{1999}.  in
  \emph{\bibinfo{booktitle}{New Approaches to Problems in Liquid State
  Theory}}, edited by
  \bibinfo{editor}{\bibfnamefont{C.}~\bibnamefont{Caccamo}},
  \bibinfo{editor}{\bibfnamefont{J.-P.} \bibnamefont{Hansen}},
  \bibinfo{editor}{\bibfnamefont{G.}~\bibnamefont{Stell}}, 
  \bibinfo{publisher}{Kluwer}, \bibinfo{address}{Netherlands},
  \bibinfo{volume}{529}:\bibinfo{pages}{407--420}

\bibitem{HummerG:Newphe}
\bibinfo{author}{\bibnamefont{Hummer} \bibfnamefont{G}},
  \bibinfo{author}{\bibnamefont{Garde} \bibfnamefont{S}},
  \bibinfo{author}{\bibnamefont{Garc\'{i}a} \bibfnamefont{A.}} \bibinfo{year}{2000}.
  \bibinfo{author}{\bibnamefont{Pratt} \bibfnamefont{LR}},
  \bibinfo{journal}{{\it Chem. Phys.\/}} \bibinfo{volume}{258}:\bibinfo{pages}{349}

\bibitem{Lee:85}
\bibinfo{author}{\bibnamefont{Lee} \bibfnamefont{B.}} \bibinfo{year}{1985}.
  \bibinfo{journal}{{\it Biopolymers\/}} \bibinfo{volume}{24}:\bibinfo{pages}{813}

\bibitem{Pohorille:JACS:90}
\bibinfo{author}{\bibnamefont{Pohorille} \bibfnamefont{A,}}
  \bibinfo{author}{\bibnamefont{Pratt} \bibfnamefont{LR.}} \bibinfo{year}{1990}.
  \bibinfo{journal}{{\it J. Am. Chem. Soc.\/}} \bibinfo{volume}{112}:\bibinfo{pages}{5066}

\bibitem{Pratt:PNAS:92}
\bibinfo{author}{\bibnamefont{Pratt} \bibfnamefont{LR,}}
  \bibinfo{author}{\bibnamefont{Pohorille} \bibfnamefont{A.}} \bibinfo{year}{1992}.
  \bibinfo{journal}{{\it Proc. Natl. Acad. Sci. USA\/}} \bibinfo{volume}{89}:\bibinfo{pages}{2995}

\bibitem{Palma}
\bibinfo{author}{\bibnamefont{Pratt} \bibfnamefont{LR,}}
  \bibinfo{author}{\bibnamefont{Pohorille},  \bibfnamefont{A.}} \bibinfo{year}{1993}. in
  \emph{\bibinfo{booktitle}{Proceedings of the {EBSA} International Workshop on
  Water-Biomolecule Interactions}}, edited by
  \bibinfo{editor}{\bibfnamefont{MU.} \bibnamefont{Palma}},
  \bibinfo{editor}{\bibfnamefont{MB.} \bibnamefont{Palma-Vittorelli}},
  \bibinfo{editor}{\bibfnamefont{F.}~\bibnamefont{Parak}}
  \bibinfo{publisher}{Societ\'{a} Italiana de Fisica},
  \bibinfo{address}{Bologna},  
  \bibinfo{pages}{261--268}

\bibitem{Chandler:PRE:93}
\bibinfo{author}{\bibnamefont{Chandler} \bibfnamefont{D.}} \bibinfo{year}{1993}.
  \bibinfo{journal}{{\it Phys. Rev. E\/}} \bibinfo{volume}{48}:\bibinfo{pages}{2898}

\bibitem{Pratt:JCP:77}
\bibinfo{author}{\bibnamefont{Pratt} \bibfnamefont{LR,}}
  \bibinfo{author}{\bibnamefont{Chandler} \bibfnamefont{D.}} \bibinfo{year}{1977}.
  \bibinfo{journal}{{\it J. Chem. Phys.\/}} \bibinfo{volume}{67}:\bibinfo{pages}{3683}

\bibitem{Chan:79}
\bibinfo{author}{\bibnamefont{Chan} \bibfnamefont{DYC}},
  \bibinfo{author}{\bibnamefont{Mitchell} \bibfnamefont{DJ}},
  \bibinfo{author}{ \bibnamefont{Ninham} \bibfnamefont{BW}},
  \bibinfo{author}{\bibnamefont{Pailthorpe} \bibfnamefont{BA.}} \bibinfo{year}{1979}.
  in \emph{\bibinfo{series}{Water: A Comprehensive Treatise}},
  edited by \bibinfo{editor}{\bibfnamefont{F.}~\bibnamefont{Franks}}
  \bibinfo{address}{New York}: \bibinfo{publisher}{Plenum},  vol~\bibinfo{volume}{6, Recent Advances}, pp.
  \bibinfo{pages}{239--278}.


\bibitem{PrattLR:Quatal}
\bibinfo{author}{\bibnamefont{Pratt} \bibfnamefont{L,}}
  \bibinfo{author}{\bibnamefont{LaViolette} \bibfnamefont{RA.}} \bibinfo{year}{1998}.
  \bibinfo{journal}{{\it Mol. Phys.\/}}  \bibinfo{volume}{94}:\bibinfo{pages}{909}

\bibitem{Pratt:ES:99}
\bibinfo{author}{\bibnamefont{Pratt} \bibfnamefont{LR,}}
  \bibinfo{author}{\bibnamefont{Rempe} \bibfnamefont{SB.}} \bibinfo{year}{1999}. in
  \emph{\bibinfo{booktitle}{Simulation and Theory of Electrostatic Interactions
  in Solution. Computational Chemistry, Biophysics, and Aqueous Solutions}},
  edited by \bibinfo{editor}{\bibfnamefont{LR.} \bibnamefont{Pratt}}
  \bibinfo{editor}{\bibfnamefont{G.}~\bibnamefont{Hummer}}
  \bibinfo{publisher}{American Institute of Physics},
  \bibinfo{address}{Melville, NY}  
   \emph{\bibinfo{series}{AIP Conference Proceedings}},
  \bibinfo{volume}{492}: \bibinfo{pages}{172--201}.

\bibitem{Rempe:JACS:2000}
\bibinfo{author}{\bibnamefont{Rempe} \bibfnamefont{SB}},
  \bibinfo{author}{\bibnamefont{Pratt} \bibfnamefont{LR}},
  \bibinfo{author}{\bibnamefont{Hummer} \bibfnamefont{G}},
  \bibinfo{author}{\bibnamefont{Kress} \bibfnamefont{JD}},
  \bibinfo{author}{\bibnamefont{Martin} \bibfnamefont{RL}},
  \bibinfo{author}{\bibnamefont{Redondo} \bibfnamefont{A.}} \bibinfo{year}{2000}.
  \bibinfo{journal}{{\it J. Am. Chem. Soc.\/}} \bibinfo{volume}{122}:\bibinfo{pages}{966} \bibinfo{year}{2000}.

\bibitem{Rempe:FPE:2001}
\bibinfo{author}{\bibnamefont{Rempe}} \bibfnamefont{SB,}
  \bibinfo{author}{\bibnamefont{Pratt} \bibfnamefont{LR.}} \bibinfo{year}{2001}.
  \bibinfo{journal}{{\it Fluid Phase Equilibria\/}}
  \bibinfo{volume}{183-184}: \bibinfo{pages}{121}
 

\bibitem{Pratt:2001}
\bibinfo{author}{\bibnamefont{Pratt} \bibfnamefont{LR}},
  \bibinfo{author}{\bibnamefont{LaViolette} \bibfnamefont{RA}},
  \bibinfo{author}{\bibnamefont{Gomez} \bibfnamefont{MA}},
  \bibinfo{author}{\bibnamefont{Gentile} \bibfnamefont{ME}}. \bibinfo{year}{2001 (in press)}.
  \bibinfo{journal}{{\it J. Phys. Chem. B\/}} \bibinfo{volume}{xxx}:yyy
  

\bibitem{Privalov:90}
\bibinfo{author}{\bibnamefont{Privalov} \bibfnamefont{PL.} } \bibinfo{year}{1990.}
  \bibinfo{journal}{{\it Crit. Rev. Biochem. Mol. Biol.\/}}
	\bibinfo{volume}{25}:\bibinfo{pages}{281}

\bibitem{ParsegianVA:Osmscp}
\bibinfo{author}{\bibnamefont{Parsegian} \bibfnamefont{V}},
  \bibinfo{author}{\bibnamefont{Rand} \bibfnamefont{R}},
  \bibinfo{author}{\bibnamefont{Rau} \bibfnamefont{D.}} \bibinfo{year}{2000}.
  \bibinfo{journal}{{\it Proc. Nat. Acad. Sci. USA\/}} \bibinfo{volume}{97}:\bibinfo{pages}{3987} 

  
\bibitem{DASHEVSKYVG:SOLHIN}
\bibinfo{author}{\bibnamefont{Dashevsky} \bibfnamefont{V}},
  \bibinfo{author}{\bibnamefont{Sarkisov} \bibfnamefont{G.}} \bibinfo{year}{1974}.
  \bibinfo{journal}{{\it Mol. Phys.\/}}  \bibinfo{volume}{27}:\bibinfo{pages}{1271}

\bibitem{OWICKIJC:MONCIE}
\bibinfo{author}{\bibnamefont{Owicki} \bibfnamefont{J}},
  \bibinfo{author}{\bibnamefont{Scheraga} \bibfnamefont{HA.}} \bibinfo{year}{1977}.
  \bibinfo{journal}{{\it J. Am. Chem. Soc.\/}} \bibinfo{volume}{99}:\bibinfo{pages}{7413}

\bibitem{SWAMINATHANS:MONSSD}
\bibinfo{author}{\bibnamefont{Swaminathan} \bibfnamefont{S}},
  \bibinfo{author}{ \bibnamefont{Harrison} \bibfnamefont{SW}},
  \bibinfo{author}{\bibnamefont{Beveridge}} \bibfnamefont{DL.} 
   \bibinfo{year}{1978}.
  \bibinfo{journal}{{\it J. Am. Chem. Soc.\/}}
  \bibinfo{volume}{100}: \bibinfo{pages}{5705}

\bibitem{GEIGERA:MOLSTH}
\bibinfo{author}{\bibnamefont{Geiger} \bibfnamefont{A}},
  \bibinfo{author}{\bibnamefont{Rahman} \bibfnamefont{A}},
  \bibinfo{author}{\bibnamefont{Stillinger, \bibfnamefont{FH.}}} \bibinfo{year}{1979}.
  \bibinfo{journal}{{\it J. Chem. Phys.\/}} \bibinfo{volume}{70}:\bibinfo{pages}{263}
  
\bibitem{PANGALIC:HYDHAP}
\bibinfo{author}{\bibnamefont{Pangali} \bibfnamefont{C}},
  \bibinfo{author}{\bibnamefont{Rao} \bibfnamefont{M}},
  \bibinfo{author}{\bibnamefont{Berne} \bibfnamefont{BJ.}} \bibinfo{year}{1979}.
  \bibinfo{journal}{{\it J. Chem. Phys.\/}} \bibinfo{volume}{71}:\bibinfo{pages}{2982}

\bibitem{BIGOTB:SAMMMS}
\bibinfo{author}{\bibnamefont{Bigot} \bibfnamefont{B}},
  \bibinfo{author}{\bibnamefont{Jorgensen} \bibfnamefont{WL.}} \bibinfo{year}{1981}.
  \bibinfo{journal}{{\it J. Chem. Phys.\/}} \bibinfo{volume}{75}:\bibinfo{pages}{1944}

\bibitem{POSTMAJPM:THECFW}
\bibinfo{author}{\bibnamefont{Postma} \bibfnamefont{J}},
  \bibinfo{author}{\bibnamefont{Berendsen} \bibfnamefont{H}},
  \bibinfo{author}{\bibnamefont{Haak} \bibfnamefont{J.}}  \bibinfo{year}{1982}.
  \bibinfo{journal}{{\it Faraday Symp. Chem. Soc.\/}} p\bibinfo{pages}{55}


\bibitem{OKAZAKIS:TEMETH}
\bibinfo{author}{\bibnamefont{Okazaki} \bibfnamefont{S}},
  \bibinfo{author}{\bibnamefont{Touhara} \bibfnamefont{H}},
  \bibinfo{author}{\bibnamefont{Nakanishi} \bibfnamefont{K}},
  \bibinfo{author}{\bibnamefont{Watanabe} \bibfnamefont{N.}} \bibinfo{year}{1982}.
  \bibinfo{journal}{{\it Bull. Chem. Soc. Jap.\/}} \bibinfo{volume}{55}:\bibinfo{pages}{2827}

\bibitem{KINCAIDRH:ACCCMS}
\bibinfo{author}{\bibnamefont{Kincaid} \bibfnamefont{R}},
  \bibinfo{author}{\bibnamefont{Scheraga} \bibfnamefont{HA.}} \bibinfo{year}{1982}.
  \bibinfo{journal}{{\it J. Comp. Chem.\/}} \bibinfo{volume}{3}:\bibinfo{pages}{525}

\bibitem{ROSSKYPJ:MOLLSO}
\bibinfo{author}{\bibnamefont{Rossky} \bibfnamefont{P}},
  \bibinfo{author}{\bibnamefont{Zichi} \bibfnamefont{DA}}.   \bibinfo{year}{1982}.
  \bibinfo{journal}{{\it Faraday Symp. Chem. Soc.\/}} p\bibinfo{pages}{69}


\bibitem{RAPAPORTDC:HYDISM}
\bibinfo{author}{\bibnamefont{Rapaport} \bibfnamefont{DC}},
  \bibinfo{author}{\bibnamefont{Scheraga} \bibfnamefont{HA.}} \bibinfo{year}{1982}.
  \bibinfo{journal}{{\it J. Phys. Chem.\/}} \bibinfo{volume}{86}:\bibinfo{pages}{873}

\bibitem{TANIA:NONSWP}
\bibinfo{author}{\bibnamefont{Tani} \bibfnamefont{A}}, \bibinfo{year}{1983}. 
\bibinfo{journal}{{\it Mol. Phys.\/}}  \bibinfo{volume}{48}: \bibinfo{pages}{1229}
  

\bibitem{SWOPEWC:MOLMCS}
\bibinfo{author}{\bibnamefont{Swope} \bibfnamefont{W}},
  \bibinfo{author}{\bibnamefont{Andersen} \bibfnamefont{HC.}} \bibinfo{year}{1984}.
  \bibinfo{journal}{{\it J. Phys. Chem.\/}} \bibinfo{volume}{88}:\bibinfo{pages}{6548}

\bibitem{Remerie:MP:84}
\bibinfo{author}{\bibnamefont{Remerie} \bibfnamefont{K}},
  \bibinfo{author}{\bibnamefont{van~Gunsteren} \bibfnamefont{WF}},
  \bibinfo{author}{\bibnamefont{Postma} \bibfnamefont{JPM}},
  \bibinfo{author}{\bibnamefont{Berendsen} \bibfnamefont{HJC}},
  \bibinfo{author}{\bibnamefont{Engberts}
  \bibfnamefont{JBFN}}.  \bibinfo{year}{1984}. \bibinfo{journal}{{\it Mol. Phys.\/}} 
  \bibinfo{volume}{53}: \bibinfo{pages}{1517}

\bibitem{LINSEP:MONSDA}
\bibinfo{author}{\bibnamefont{Linse} \bibfnamefont{P}},
  \bibinfo{author}{\bibnamefont{Karlstrom} \bibfnamefont{G}},
  \bibinfo{author}{\bibnamefont{Jonsson} \bibfnamefont{B.}} \bibinfo{year}{1984}.
  \bibinfo{journal}{{\it J. Am. Chem. Soc.\/}} \bibinfo{volume}{106}:\bibinfo{pages}{4096} 

\bibitem{JORGENSENWL:MONSAW}
\bibinfo{author}{\bibnamefont{Jorgensen} \bibfnamefont{WL}},
  \bibinfo{author}{\bibnamefont{Gao} \bibfnamefont{J}},
  \bibinfo{author}{\bibnamefont{Ravimohan} \bibfnamefont{C.}} \bibinfo{year}{1985}.
  \bibinfo{journal}{{\it J. Phys. Chem.\/}} \bibinfo{volume}{89}:\bibinfo{pages}{3470}

\bibitem{STRAATSMATP:FREHHM}
\bibinfo{author}{\bibnamefont{Straatsma} \bibfnamefont{TP}},
  \bibinfo{author}{\bibnamefont{Berendsen} \bibfnamefont{HJC}},
  \bibinfo{author}{\bibnamefont{Postma},
  \bibfnamefont{JPM.}} \bibinfo{year}{1986}. 
  \bibinfo{journal}{{\it J. Chem. Phys.\/}}
  \bibinfo{volume}{85}: \bibinfo{pages}{6720}

\bibitem{ZICHIDA:SOLMRH}
\bibinfo{author}{\bibnamefont{Zichi} \bibfnamefont{DA}},
  \bibinfo{author}{\bibnamefont{Rossky} \bibfnamefont{P.}} \bibinfo{year}{1986}.
  \bibinfo{journal}{{\it J. Chem. Phys.\/}} \bibinfo{volume}{84}:\bibinfo{pages}{2814}
  
\bibitem{ZICHIDA:MOLCEL}
\bibinfo{author}{\bibnamefont{Zichi} \bibfnamefont{DA}},
  \bibinfo{author}{\bibnamefont{Rossky} \bibfnamefont{P.}} \bibinfo{year}{1986}.
  \bibinfo{journal}{{\it J. Chem. Phys.\/}} \bibinfo{volume}{84}:\bibinfo{pages}{1712}

\bibitem{FOISES:MONSAN}
\bibinfo{author}{\bibnamefont{Fois} \bibfnamefont{E}},
  \bibinfo{author}{\bibnamefont{Gamba} \bibfnamefont{A}},
  \bibinfo{author}{\bibnamefont{Morosi} \bibfnamefont{G}},
  \bibinfo{author}{\bibnamefont{Demontis} \bibfnamefont{P}},
  \bibinfo{author}{\bibnamefont{Suffritti} \bibfnamefont{G.}} \bibinfo{year}{1986}.
  \bibinfo{journal}{{\it Mol. Phys.\/}}  \bibinfo{volume}{58}:\bibinfo{pages}{65}

\bibitem{TANAKAH:INTMSH}
\bibinfo{author}{\bibnamefont{Tanaka} \bibfnamefont{H.}} \bibinfo{year}{1987}. 
\bibinfo{journal}{{\it J. Chem. Phys.\/}} \bibinfo{volume}{86}: \bibinfo{pages}{1512}
 

\bibitem{FLEISCHMANSH:THEASS}
\bibinfo{author}{\bibnamefont{Fleischman} \bibfnamefont{SH,}}
  \bibinfo{author}{\bibnamefont{Brooks} \bibfnamefont{CL.}} \bibinfo{year}{1987}. 
  \bibinfo{journal}{{\it J. Chem. Phys.\/}}
  \bibinfo{volume}{87}: \bibinfo{pages}{3029}

\bibitem{KOOPOY:MODSHS}
\bibinfo{author}{\bibnamefont{Koop} \bibfnamefont{O,}}
  \bibinfo{author}{\bibnamefont{Perelygin} \bibinfo{year}{1988}. \bibfnamefont{I}},
  \bibinfo{journal}{{\it Z. Fiz. Khim. (English Tranlation)\/}}
  \bibinfo{volume}{62}: \bibinfo{pages}{1070}

\bibitem{JORGENSENWL:FRETWF}
\bibinfo{author}{ \bibnamefont{Jorgensen} \bibfnamefont{WL}},
  \bibinfo{author}{\bibnamefont{Blake} \bibfnamefont{JF}},
  \bibinfo{author}{\bibnamefont{Buckner} \bibfnamefont{JK.}} \bibinfo{year}{1989}.
  \bibinfo{journal}{{\it Chem. Phys.\/}} \bibinfo{volume}{129}:\bibinfo{pages}{193}

\bibitem{RAOBG:HYDHFP}
\bibinfo{author}{\bibnamefont{Rao} \bibfnamefont{BG,}}
  \bibinfo{author}{\bibnamefont{Singh} \bibfnamefont{UC.}} \bibinfo{year}{1989}.
  \bibinfo{journal}{{\it J. Am. Chem. Soc.\/}} \bibinfo{volume}{111}:\bibinfo{pages}{3125}

\bibitem{LINSEP:MOLSDA}
\bibinfo{author}{\bibnamefont{Linse} \bibfnamefont{P.}} \bibinfo{year}{1990}. 
\bibinfo{journal}{{\it J. Am. Chem. Soc.\/}} \bibinfo{volume}{112}: \bibinfo{pages}{1744}
 

\bibitem{GUILLOTB:COMSHH}
\bibinfo{author}{\bibnamefont{Guillot} \bibfnamefont{B}},
  \bibinfo{author}{\bibnamefont{Guissani} \bibfnamefont{Y}},
  \bibinfo{author}{\bibnamefont{Bratos} \bibfnamefont{S.}} \bibinfo{year}{1991}.
  \bibinfo{journal}{{\it J. Chem. Phys.\/}} \bibinfo{volume}{95}:\bibinfo{pages}{3643}

\bibitem{LAAKSONENA:MOLNMW}
\bibinfo{author}{\bibnamefont{Laaksonen} \bibfnamefont{A,}}
  \bibinfo{author}{\bibnamefont{Stilbs} \bibfnamefont{P.}} \bibinfo{year}{1991}.
  \bibinfo{journal}{{\it Mol. Phys.\/}}  \bibinfo{volume}{74}:\bibinfo{pages}{747}

\bibitem{CUMMINGSPT:SIMSWS}
\bibinfo{author}{\bibnamefont{Cummings} \bibfnamefont{PT}},
  \bibinfo{author}{\bibnamefont{Cochran} \bibfnamefont{HD}},
  \bibinfo{author}{\bibnamefont{Simonson} \bibfnamefont{JM}},
  \bibinfo{author}{\bibnamefont{Mesmer} \bibfnamefont{RE}},
  \bibinfo{author}{\bibnamefont{Karaborni} \bibfnamefont{S.}} \bibinfo{year}{1991}.
  \bibinfo{journal}{{\it J. Chem. Phys.\/}} \bibinfo{volume}{94}:\bibinfo{pages}{5606}

\bibitem{TANAKAH:HYDHIT}
\bibinfo{author}{\bibnamefont{Tanaka} \bibfnamefont{H,}}
  \bibinfo{author}{\bibnamefont{Nakanishi} \bibfnamefont{K.}} \bibinfo{year}{1991}.
  \bibinfo{journal}{{\it J. Chem. Phys.\/}} \bibinfo{volume}{95}:\bibinfo{pages}{3719}

\bibitem{andaloro}
\bibinfo{author}{\bibnamefont{Andaloro} \bibfnamefont{G,}}
  \bibinfo{author}{\bibnamefont{Sperandeo-Mineo} \bibfnamefont{RM.}} \bibinfo{year}{1991}.
  \bibinfo{journal}{{\it Eur. J. Phys.\/}} \bibinfo{volume}{11}:\bibinfo{pages}{275}

\bibitem{FLEISCHMANSH:FRESMS}
\bibinfo{author}{\bibnamefont{Fleischman} \bibfnamefont{SH,}}
  \bibinfo{author}{\bibnamefont{Zichi} \bibfnamefont{DA}}, \bibinfo{year}{1991}.
  \bibinfo{journal}{{\it J. Chim. Phys. Phys.-Chim. Bio.\/}}
  \bibinfo{volume}{88}: \bibinfo{pages}{2617}

\bibitem{Lazaridis:92}
\bibinfo{author}{\bibnamefont{Lazaridis} \bibfnamefont{T,}}
  \bibinfo{author}{\bibnamefont{Paulaitis} \bibfnamefont{ME.}} \bibinfo{year}{1992}.
  \bibinfo{journal}{{\it J. Phys. Chem.\/}} \bibinfo{volume}{96}:\bibinfo{pages}{3847}

\bibitem{Wallqvist:JCP:92}
\bibinfo{author}{\bibnamefont{Wallqvist} \bibfnamefont{A.}} \bibinfo{year}{1992}.
  \bibinfo{journal}{{\it J. Chem. Phys.\/}} \bibinfo{volume}{96}:\bibinfo{pages}{1655}

\bibitem{SUNYX:SIMTSF}
\bibinfo{author}{\bibnamefont{Sun} \bibfnamefont{Y}},
  \bibinfo{author}{\bibnamefont{Spellmeyer} \bibfnamefont{D}},
  \bibinfo{author}{\bibnamefont{Pearlman} \bibfnamefont{D}},
  \bibinfo{author}{\bibnamefont{Kollman} \bibfnamefont{P.}} \bibinfo{year}{1992}.
  \bibinfo{journal}{{\it J. Am. Chem. Soc.\/}} \bibinfo{volume}{114}:\bibinfo{pages}{6798} 

\bibitem{LAZARIDIST:ENTHHN}
\bibinfo{author}{\bibnamefont{Lazaridis} \bibfnamefont{T,}}
  \bibinfo{author}{\bibnamefont{Paulaitis} \bibfnamefont{ME.}}\bibinfo{year}{1992}.
  \bibinfo{journal}{{\it J. Phys. Chem.\/}} \bibinfo{volume}{96}:\bibinfo{pages}{3847} 

\bibitem{Guillot:JCP:93}
\bibinfo{author}{\bibnamefont{Guillot} \bibfnamefont{B,}}
  \bibinfo{author}{\bibnamefont{Guissani} \bibfnamefont{Y.}} \bibinfo{year}{1993}.
  \bibinfo{journal}{{\it J. Chem. Phys.\/}} \bibinfo{volume}{99}:\bibinfo{pages}{8075}

\bibitem{GIIOB:AQUNGC}
\bibinfo{author}{\bibnamefont{Guillot} \bibfnamefont{B}},
  \bibinfo{author}{\bibnamefont{Guissanni} \bibfnamefont{Y}},
  \bibinfo{author}{\bibnamefont{Bratos} \bibfnamefont{S.}} \bibinfo{year}{1993}.
  \bibinfo{journal}{{\it Z. Fiz. Khim. (English Tranlation)\/}}
  \bibinfo{volume}{67}: \bibinfo{pages}{25}

\bibitem{SMITHDE:FREEIE}
\bibinfo{author}{\bibnamefont{Smith} \bibfnamefont{D,}}
  \bibinfo{author}{\bibnamefont{Haymet} \bibfnamefont{A.}} \bibinfo{year}{1993}.
  \bibinfo{journal}{{\it J. Chem. Phys.\/}} \bibinfo{volume}{98}:\bibinfo{pages}{6445}

\bibitem{VANBELLED:MOLSMH}
\bibinfo{author}{\bibnamefont{van Belle} \bibfnamefont{D,}}
  \bibinfo{author}{\bibnamefont{Wodak} \bibfnamefont{SJ.}} \bibinfo{year}{1993}.
  \bibinfo{journal}{{\it J. Am. Chem. Soc.\/}} \bibinfo{volume}{115}:\bibinfo{pages}{647}

\bibitem{GUILLOTB:TEMTSN}
\bibinfo{author}{\bibnamefont{Guillot} \bibfnamefont{B,}}
  \bibinfo{author}{\bibnamefont{Guissani} \bibfnamefont{Y.}} \bibinfo{year}{1993}.
  \bibinfo{journal}{{\it Mol. Phys.\/}}  \bibinfo{volume}{79}:\bibinfo{pages}{53}

\bibitem{ZENGJ:ENTHPM}
\bibinfo{author}{\bibnamefont{Zeng} \bibfnamefont{J}},
  \bibinfo{author}{\bibnamefont{Hush} \bibfnamefont{NS}},
  \bibinfo{author}{\bibnamefont{Reimers} \bibfnamefont{JR.}} \bibinfo{year}{1993}.
  \bibinfo{journal}{{\it Chem. Phys. Letts.\/}} \bibinfo{volume}{206}:\bibinfo{pages}{318} 

\bibitem{SKIPPERNT:COMMWS}
\bibinfo{author}{\bibnamefont{Skipper} \bibfnamefont{NT.}} \bibinfo{year}{1993}.
  \bibinfo{journal}{{\it Chem. Phys. Letts.\/}} \bibinfo{volume}{207}:\bibinfo{pages}{424}

\bibitem{Beglov:94}
\bibinfo{author}{\bibnamefont{Beglov} \bibfnamefont{D,}}
  \bibinfo{author}{\bibnamefont{Roux} \bibfnamefont{B.}} \bibinfo{year}{1994}. 
  \bibinfo{journal}{{\it J. Chem. Phys.\/}} \bibinfo{volume}{100}: \bibinfo{pages}{9050}
 

\bibitem{FORSMANJ:MONSHI}
\bibinfo{author}{\bibnamefont{Forsman} \bibfnamefont{J,}}
  \bibinfo{author}{\bibnamefont{Jonsson} \bibfnamefont{B.}} \bibinfo{year}{1994}.
  \bibinfo{journal}{{\it J. Chem. Phys.\/}} \bibinfo{volume}{101}:\bibinfo{pages}{5116}

\bibitem{MADANB:ROLHHF}
\bibinfo{author}{\bibnamefont{Madan} \bibfnamefont{B,}}
  \bibinfo{author}{\bibnamefont{Lee} \bibfnamefont{B.}} \bibinfo{year}{1994}.
  \bibinfo{journal}{{\it Biophys. Chem.\/}} \bibinfo{volume}{51}:\bibinfo{pages}{279}

\bibitem{MATUBAYASIN:MATWGH}
\bibinfo{author}{\bibnamefont{Matubayasi} \bibfnamefont{N.}} \bibinfo{year}{1994}.
  \bibinfo{journal}{{\it J. Am. Chem. Soc.\/}} \bibinfo{volume}{116}:\bibinfo{pages}{1450}

\bibitem{MATUBAYASIN:THETHS:1}
\bibinfo{author}{\bibnamefont{Matubayasi} \bibfnamefont{N}},
  \bibinfo{author}{\bibnamefont{Reed} \bibfnamefont{L}},
  \bibinfo{author}{\bibnamefont{Levy} \bibfnamefont{RM.}} \bibinfo{year}{1994}.
  \bibinfo{journal}{{\it J. Phys. Chem.\/}} \bibinfo{volume}{98}:\bibinfo{pages}{10640}

\bibitem{BUSHUEVYG:STRCHS}
\bibinfo{author}{\bibnamefont{Bushuev} \bibfnamefont{Y.}} \bibinfo{year}{1994}.
  \bibinfo{journal}{{\it Zhurnal Obshchei Khimii\/}} \bibinfo{volume}{64}:\bibinfo{pages}{1931} 

\bibitem{LAZARIDIST:SIMSTH}
\bibinfo{author}{\bibnamefont{Lazaridis} \bibfnamefont{T,}}
  \bibinfo{author}{\bibnamefont{Paulaitis} \bibfnamefont{ME.}} \bibinfo{year}{1994}.
  \bibinfo{journal}{{\it J. Phys. Chem.\/}} \bibinfo{volume}{98}:\bibinfo{pages}{635}

\bibitem{KAMINSKIG:FREHPL}
\bibinfo{author}{\bibnamefont{Kaminski} \bibfnamefont{G}},
  \bibinfo{author}{\bibnamefont{Duffy} \bibfnamefont{EM}},
  \bibinfo{author}{\bibnamefont{Matsui} \bibfnamefont{T}},
  \bibinfo{author}{\bibnamefont{Jorgensen} \bibfnamefont{WL.}} \bibinfo{year}{1994}.
  \bibinfo{journal}{{\it J. Phys. Chem.\/}} \bibinfo{volume}{98}:\bibinfo{pages}{13077}

\bibitem{headgordont:effsms}
\bibinfo{author}{\bibnamefont{Head-Gordon} \bibfnamefont{T.}} \bibinfo{year}{1994}.
  \bibinfo{journal}{{\it Chem. Phys. Letts.\/}} \bibinfo{volume}{227}:\bibinfo{pages}{215}

\bibitem{BEUTLERTC:FRECFS}
\bibinfo{author}{\bibnamefont{Beutler} \bibfnamefont{TC}},
  \bibinfo{author}{\bibnamefont{Beguelin} \bibfnamefont{DR}},
  \bibinfo{author}{\bibnamefont{van~Gunsteren}
  \bibfnamefont{WF.}} \bibinfo{year}{1995}.
  \bibinfo{journal}{{\it J. Chem. Phys.\/}}
  \bibinfo{volume}{102}: \bibinfo{pages}{3787}

\bibitem{Head-Gordon:JACS:95}
\bibinfo{author}{\bibnamefont{Head-Gordon} \bibfnamefont{T.}} \bibinfo{year}{1995}.
  \bibinfo{journal}{{\it J. Am. Chem. Soc.\/}} \bibinfo{volume}{117}:\bibinfo{pages}{501}

\bibitem{SUNYX:HYDSMN}
\bibinfo{author}{ \bibnamefont{Sun} \bibfnamefont{YX,}}
  \bibinfo{author}{\bibnamefont{Kollman} \bibfnamefont{PA.}} \bibinfo{year}{1995}.
  \bibinfo{journal}{{\it J. Comp. Chem.\/}} \bibinfo{volume}{16}:\bibinfo{pages}{1164}

\bibitem{Wallqvist:JPC:95:a}
\bibinfo{author}{\bibnamefont{Wallqvist} \bibfnamefont{A,}}
  \bibinfo{author}{\bibnamefont{Berne} \bibfnamefont{BJ.}} \bibinfo{year}{1995}.
  \bibinfo{journal}{{\it J. Phys. Chem.\/}} \bibinfo{volume}{99}:\bibinfo{pages}{2885}

\bibitem{Head-Gordon:PNAS:95}
\bibinfo{author}{\bibnamefont{Head-Gordon} \bibfnamefont{T.}} \bibinfo{year}{1995}.
  \bibinfo{journal}{{\it Proc. Natl. Acad. Sci. USA\/}} \bibinfo{volume}{92}:\bibinfo{pages}{8308}

\bibitem{MANCERARL:TEMETH}
\bibinfo{author}{\bibnamefont{Mancera} \bibfnamefont{RL,}}
  \bibinfo{author}{\bibnamefont{Buckingham} \bibfnamefont{AD.}} \bibinfo{year}{1995}.
  \bibinfo{journal}{{\it J. Phys. Chem.\/}} \bibinfo{volume}{99}:\bibinfo{pages}{14632}

\bibitem{AshbaughHS:EnthhE}
\bibinfo{author}{ \bibnamefont{Ashbaugh} \bibfnamefont{HS,}}
  \bibinfo{author}{\bibnamefont{Paulaitis} \bibfnamefont{ME.}} \bibinfo{year}{1996}.
  \bibinfo{journal}{{\it J. Phys. Chem.\/}} \bibinfo{volume}{100}:\bibinfo{pages}{1900}

\bibitem{Garde:PRE:96}
\bibinfo{author}{\bibnamefont{Garde} \bibfnamefont{S}},
  \bibinfo{author}{\bibnamefont{Hummer} \bibfnamefont{G}},
  \bibinfo{author}{\bibnamefont{{Garc\'{\i}a}} \bibfnamefont{AE}},
  \bibinfo{author}{\bibnamefont{Pratt} \bibfnamefont{LR}},
  \bibinfo{author}{\bibnamefont{Paulaitis} \bibfnamefont{ME.}} \bibinfo{year}{1996}.
  \bibinfo{journal}{{\it Phys. Rev. E\/}} \bibinfo{volume}{53}:\bibinfo{pages}{R4310}

\bibitem{Prevost:JPC:96}
\bibinfo{author}{\bibnamefont{Prevost} \bibfnamefont{M}},
  \bibinfo{author}{\bibnamefont{Oliveira} \bibfnamefont{IT}},
  \bibinfo{author}{\bibnamefont{Kocher} \bibfnamefont{JP}},
  \bibinfo{author}{\bibnamefont{Wodak} \bibfnamefont{SJ.}} \bibinfo{year}{1996}.
  \bibinfo{journal}{{\it J. Phys. Chem.\/}} \bibinfo{volume}{100}:\bibinfo{pages}{2738}

\bibitem{ChauPL:Curehs}
\bibinfo{author}{\bibnamefont{Chau} \bibfnamefont{P}},
  \bibinfo{author}{\bibnamefont{Forester} \bibfnamefont{T}},
  \bibinfo{author}{\bibnamefont{Smith} \bibfnamefont{W.}} \bibinfo{year}{1996}.
  \bibinfo{journal}{{\it Mol. Phys.\/}}  \bibinfo{volume}{89}:\bibinfo{pages}{1033}

\bibitem{ReM:Strctd}
\bibinfo{author}{\bibnamefont{Re} \bibfnamefont{M}},
  \bibinfo{author}{\bibnamefont{Laria} \bibfnamefont{D}},
  \bibinfo{author}{\bibnamefont{Fern\'{a}ndez-Prini} \bibfnamefont{R.}} \bibinfo{year}{1996}.
  \bibinfo{journal}{{\it Chem. Phys. Letts.\/}} \bibinfo{volume}{250}:\bibinfo{pages}{25}

\bibitem{MatubayasiN:Theths:2}
\bibinfo{author}{\bibnamefont{Matubayasi} \bibfnamefont{N,}}
  \bibinfo{author}{\bibnamefont{Levy} \bibfnamefont{RM.}} \bibinfo{year}{1996}.
  \bibinfo{journal}{{\it J. Phys. Chem.\/}} \bibinfo{volume}{100}:\bibinfo{pages}{2681}

\bibitem{DurellSR:Atoats}
\bibinfo{author}{\bibnamefont{Durell} \bibfnamefont{SR,}}
  \bibinfo{author}{\bibnamefont{Wallqvist} \bibfnamefont{A.}} \bibinfo{year}{1996}.
  \bibinfo{journal}{{\it Biophys. J.\/}} \bibinfo{volume}{71}:\bibinfo{pages}{1695}

\bibitem{WallqvistA:theoth}
\bibinfo{author}{\bibnamefont{Wallqvist} \bibfnamefont{A,}}
  \bibinfo{author}{\bibnamefont{Covell} \bibfnamefont{DG.}} \bibinfo{year}{1996}.
  \bibinfo{journal}{{\it Biophys. J.\/}} \bibinfo{volume}{71}:\bibinfo{pages}{600}

\bibitem{SkipperNT:Comsst}
\bibinfo{author}{\bibnamefont{Skipper} \bibfnamefont{NT}},
  \bibinfo{author}{\bibnamefont{Bridgeman} \bibfnamefont{CH}},
  \bibinfo{author}{\bibnamefont{Buckingham} \bibfnamefont{AD}},
  \bibinfo{author}{
  \bibnamefont{Mancera} \bibfnamefont{RL.}} \bibinfo{year}{1996}.
  \bibinfo{journal}{{\it Faraday Disc.\/}} \bibinfo{volume}{103}:   \bibinfo{pages}{141--150}

\bibitem{HaymetADJ:Hydrmt}
\bibinfo{author}{\bibnamefont{Haymet} \bibfnamefont{A}},
  \bibinfo{author}{\bibnamefont{Silverstein} \bibfnamefont{K}},
  \bibinfo{author}{\bibnamefont{Dill} \bibfnamefont{K.}} \bibinfo{year}{1996}.
  \bibinfo{journal}{{\it Faraday Disc.\/}} \bibinfo{volume}{103}: 
  \bibinfo{pages}{117}

\bibitem{GardeS:HydiCe}
\bibinfo{author}{\bibnamefont{Garde} \bibfnamefont{S}},
  \bibinfo{author}{\bibnamefont{Hummer} \bibfnamefont{G}},
  \bibinfo{author}{\bibnamefont{Paulaitis} \bibfnamefont{ME.}} \bibinfo{year}{1996}.
  \bibinfo{journal}{{\it Faraday Disc.\/}} \bibinfo{volume}{103}:\bibinfo{pages}{125}

\bibitem{LinCL:Pretfe}
\bibinfo{author}{\bibnamefont{Lin} \bibfnamefont{CL,}}
  \bibinfo{author}{\bibnamefont{Wood} \bibfnamefont{RH.}} \bibinfo{year}{1996}.
  \bibinfo{journal}{{\it J. Phys. Chem.\/}} \bibinfo{volume}{100}:\bibinfo{pages}{16399}

\bibitem{ManceraRL:Hydbth}
\bibinfo{author}{\bibnamefont{Mancera} \bibfnamefont{RL.}} \bibinfo{year}{1996}.
  \bibinfo{journal}{{\it J. Chem. Soc. - Faraday Trans.\/}}
  \bibinfo{volume}{92}: \bibinfo{pages}{2547}

\bibitem{MengEC:Moldst}
\bibinfo{author}{\bibnamefont{Meng} \bibfnamefont{EC,}}
  \bibinfo{author}{\bibnamefont{Kollman} \bibfnamefont{PA.}}\bibinfo{year}{1996}.
  \bibinfo{journal}{{\it J. Phys. Chem.\/}} \bibinfo{volume}{100}:\bibinfo{pages}{11460} 

\bibitem{LyndenBellRM:hydhbs}
\bibinfo{author}{\bibnamefont{Lynden-Bell} \bibfnamefont{R,}}
  \bibinfo{author}{\bibnamefont{Rasaiah} \bibfnamefont{J.}} \bibinfo{year}{1997}.
  \bibinfo{journal}{{\it J. Chem. Phys.\/}} \bibinfo{volume}{107}:\bibinfo{pages}{1981}

\bibitem{FlorisFM:Freeei}
\bibinfo{author}{\bibnamefont{Floris} \bibfnamefont{F}},
  \bibinfo{author}{\bibnamefont{Selmi} \bibfnamefont{M}},
  \bibinfo{author}{\bibnamefont{Tani} \bibfnamefont{A}},
  \bibinfo{author}{\bibnamefont{Tomasi} \bibfnamefont{J.}} \bibinfo{year}{1997}.
  \bibinfo{journal}{{\it J. Chem. Phys.\/}} \bibinfo{volume}{107}:\bibinfo{pages}{6353}

\bibitem{DeJongPHK:Hydhm}
\bibinfo{author}{\bibnamefont{DeJong} \bibfnamefont{PHK}},
  \bibinfo{author}{\bibnamefont{Wilson} \bibfnamefont{JE}},
  \bibinfo{author}{\bibnamefont{Neilson} \bibfnamefont{GW}},
  \bibinfo{author}{
  \bibnamefont{Buckingham} \bibfnamefont{AD.}} \bibinfo{year}{1997}.
 \bibinfo{journal}{{\it Mol. Phys.\/}} 
  \bibinfo{volume}{91}: \bibinfo{pages}{99}
  
\bibitem{RadmerRJ:Freecm}
\bibinfo{author}{\bibnamefont{Radmer} \bibfnamefont{RJ,}}
  \bibinfo{author}{\bibnamefont{Kollman} \bibfnamefont{PA.}} \bibinfo{year}{1997}.
  \bibinfo{journal}{{\it J. Comp. Chem.\/}} \bibinfo{volume}{18}:\bibinfo{pages}{902}

\bibitem{mancerarl:aggmas}
\bibinfo{author}{\bibnamefont{Mancera} \bibfnamefont{RL}},
  \bibinfo{author}{\bibnamefont{Buckingham} \bibfnamefont{AD}},
  \bibinfo{author}{\bibnamefont{Skipper}
  \bibfnamefont{NT.}} \bibinfo{year}{1997}. \bibinfo{journal}{{\it J. Chem.
  Soc. - Faraday Trans.\/}}
  \bibinfo{volume}{93}: \bibinfo{pages}{2263}

\bibitem{Silverstein:98a}
\bibinfo{author}{\bibnamefont{Silverstein} \bibfnamefont{KAT}},
  \bibinfo{author}{\bibnamefont{Dill} \bibfnamefont{KA}},
  \bibinfo{author}{\bibnamefont{Haymet} \bibfnamefont{ADJ.}} \bibinfo{year}{1998}.
  \bibinfo{journal}{{\it Fluid Phase Equilibria\/}} \bibinfo{volume}{151}:\bibinfo{pages}{83}
  
\bibitem{SilversteinKAT:simmwh}
\bibinfo{author}{\bibnamefont{Silverstein} \bibfnamefont{K}},
  \bibinfo{author}{\bibnamefont{Haymet} \bibfnamefont{A}},
  \bibinfo{author}{\bibnamefont{Dill} \bibfnamefont{K.}} \bibinfo{year}{1998}. 
  \bibinfo{journal}{{\it J. Am. Chem. Soc.\/}} \bibinfo{volume}{120}: \bibinfo{pages}{3166}
 

\bibitem{IkeguchiM:Rolhbh}
\bibinfo{author}{\bibnamefont{Ikeguchi} \bibfnamefont{M}},
  \bibinfo{author}{\bibnamefont{Shimizu} \bibfnamefont{S}},
  \bibinfo{author}{\bibnamefont{Nakamura} \bibfnamefont{S}},
  \bibinfo{author}{\bibnamefont{Shimizu} \bibfnamefont{K.}} \bibinfo{year}{1998}.
  \bibinfo{journal}{{\it J. Phys. Chem. B\/}} \bibinfo{volume}{102}:\bibinfo{pages}{5891}

\bibitem{ArthurJW:Solnsw}
\bibinfo{author}{\bibnamefont{Arthur} \bibfnamefont{J,}}
  \bibinfo{author}{\bibnamefont{Haymet} \bibfnamefont{A.}} \bibinfo{year}{1998}.
  \bibinfo{journal}{{\it J. Chem. Phys.\/}} \bibinfo{volume}{109}:\bibinfo{pages}{7991}

\bibitem{PanhuisMIH:moldsc}
\bibinfo{author}{\bibnamefont{Panhuis} \bibfnamefont{M}},
  \bibinfo{author}{\bibnamefont{Patterson} \bibfnamefont{C}},
  \bibinfo{author}{\bibnamefont{Lynden-Bell} \bibfnamefont{R.}} \bibinfo{year}{1998}.
  \bibinfo{journal}{{\it Mol. Phys.\/}}  \bibinfo{volume}{94}:\bibinfo{pages}{963}

\bibitem{MountainRD:Hydsh}
\bibinfo{author}{\bibnamefont{Mountain} \bibfnamefont{RD,}}
  \bibinfo{author}{\bibnamefont{Thirumalai} \bibfnamefont{D.}} \bibinfo{year}{1998}.
  \bibinfo{journal}{{\it Proc. Nat. Acad. Sci. USA\/}} \bibinfo{volume}{95}:\bibinfo{pages}{8436}

\bibitem{ErringtonJR:Molspe}
\bibinfo{author}{\bibnamefont{Errington, \bibfnamefont{JR}}},
  \bibinfo{author}{\bibnamefont{Boulougouris} \bibfnamefont{GC}},
  \bibinfo{author}{\bibnamefont{Economou} \bibfnamefont{IG}},
  \bibinfo{author}{\bibnamefont{Panagiotopoulos} \bibfnamefont{AZ}},
  \bibinfo{author}{\bibnamefont{Theodorou},
  \bibfnamefont{DN.}} \bibinfo{year}{1998}.
  \bibinfo{journal}{{\it J. Phys. Chem. B\/}}
  \bibinfo{volume}{102}: \bibinfo{pages}{8865}
  
\bibitem{mancerarl:comste}
\bibinfo{author}{\bibnamefont{Mancera} \bibfnamefont{RL.}} \bibinfo{year}{1998}.
  \bibinfo{journal}{{\it J. Chem. Soc. - Faraday Trans.\/}}
  \bibinfo{volume}{94}: \bibinfo{pages}{3549}

\bibitem{Silverstein:99}
\bibinfo{author}{\bibnamefont{Silverstein} \bibfnamefont{KAT}},
  \bibinfo{author}{\bibnamefont{Haymet} \bibfnamefont{ADJ}},
  \bibinfo{author}{\bibnamefont{Dill} \bibfnamefont{KA.}} \bibinfo{year}{1999}.
  \bibinfo{journal}{{\it J. Chem. Phys.\/}} \bibinfo{volume}{111}:\bibinfo{pages}{8000}

\bibitem{TomasOliveiraI:Thecfw}
\bibinfo{author}{\bibnamefont{Tomas-Oliveira} \bibfnamefont{I,}}
  \bibinfo{author}{\bibnamefont{Wodak} \bibfnamefont{SJ.}} \bibinfo{year}{1999}.
  \bibinfo{journal}{{\it J. Chem. Phys.\/}} \bibinfo{volume}{111}:\bibinfo{pages}{8576}

\bibitem{Arthur:JCP:99}
\bibinfo{author}{\bibnamefont{Arthur} \bibfnamefont{JW,}}
  \bibinfo{author}{\bibnamefont{Haymet} \bibfnamefont{ADJ.}} \bibinfo{year}{1999}.
  \bibinfo{journal}{{\it J. Chem. Phys.\/}} \bibinfo{volume}{110}:\bibinfo{pages}{5873}

\bibitem{PomesR:Calecp}
\bibinfo{author}{\bibnamefont{Pomes} \bibfnamefont{R}},
  \bibinfo{author}{\bibnamefont{Eisenmesser} \bibfnamefont{E}},
  \bibinfo{author}{\bibnamefont{Post} \bibfnamefont{C}},
  \bibinfo{author}{\bibnamefont{Roux} \bibfnamefont{B.}} \bibinfo{year}{1999}. 
  \bibinfo{journal}{{\it J.
  Chem. Phys.\/}} \bibinfo{volume}{111}: \bibinfo{pages}{3387}
 

\bibitem{UrahataS:MonCst}
\bibinfo{author}{\bibnamefont{Urahata} \bibfnamefont{S,}}
  \bibinfo{author}{\bibnamefont{Canuto} \bibfnamefont{S.}} \bibinfo{year}{1999}.
  \bibinfo{journal}{{\it Chem. Phys. Letts.\/}} \bibinfo{volume}{313}:\bibinfo{pages}{235}

\bibitem{FoisE:Hydecs}
\bibinfo{author}{\bibnamefont{Fois} \bibfnamefont{E}},
  \bibinfo{author}{\bibnamefont{Gamba} \bibfnamefont{A}},
  \bibinfo{author}{\bibnamefont{Redaelli} \bibfnamefont{C.}} \bibinfo{year}{1999}.
  \bibinfo{journal}{{\it J. Chem. Phys.\/}} \bibinfo{volume}{110}:\bibinfo{pages}{1025}

\bibitem{SlusherJT:Acceic}
\bibinfo{author}{\bibnamefont{Slusher} \bibfnamefont{J.}} \bibinfo{year}{1999}. \bibinfo{journal}{{\it J.
  Phys. Chem. B\/}} \bibinfo{volume}{103}: \bibinfo{pages}{6075}
 

\bibitem{SomasundaramT:pasgtl}
\bibinfo{author}{\bibnamefont{Somasundaram} \bibfnamefont{T}},
  \bibinfo{author}{\bibnamefont{Lynden-Bell} \bibfnamefont{R}},
  \bibinfo{author}{\bibnamefont{Patterson} \bibfnamefont{C.}} \bibinfo{year}{1999}.
  \bibinfo{journal}{{\it Phys. Chem. Chem. Phys.\/}} \bibinfo{volume}{1}:\bibinfo{pages}{143}

\bibitem{ChauPL:Comsts}
\bibinfo{author}{\bibnamefont{Chau} \bibfnamefont{P,}}
  \bibinfo{author}{\bibnamefont{Mancera} \bibfnamefont{RL.}} \bibinfo{year}{1999}.
  \bibinfo{journal}{{\it Mol. Phys.\/}}  \bibinfo{volume}{96}:\bibinfo{pages}{109}

\bibitem{SmithPE:Comsce}
\bibinfo{author}{\bibnamefont{Smith} \bibfnamefont{P.}}  \bibinfo{year}{1999}. \bibinfo{journal}{{\it J.
  Phys. Chem. B\/}} \bibinfo{volume}{103}: \bibinfo{pages}{525}


\bibitem{MadanB:Chawsi}
\bibinfo{author}{\bibnamefont{Madan} \bibfnamefont{B,}}
  \bibinfo{author}{\bibnamefont{Sharp} \bibfnamefont{K.}} \bibinfo{year}{1999}.
  \bibinfo{journal}{{\it Biophys. Chem.\/}} \bibinfo{volume}{78}:\bibinfo{pages}{33}

\bibitem{GuisoniN:Squwas}
\bibinfo{author}{\bibnamefont{Guisoni}} \bibfnamefont{N,}
  \bibinfo{author}{\bibnamefont{Henriques} \bibfnamefont{V.}} \bibinfo{year}{2000}.
  \bibinfo{journal}{{\it Braz. J. Phys.\/}} \bibinfo{volume}{30}:\bibinfo{pages}{736}

\bibitem{SvishchevIM:Solstd}
\bibinfo{author}{\bibnamefont{Svishchev} \bibfnamefont{I}},
  \bibinfo{author}{\bibnamefont{Zassetsky} \bibfnamefont{A}},
  \bibinfo{author}{\bibnamefont{Kusalik} \bibfnamefont{P.}~} \bibinfo{year}{2000}.
  \bibinfo{journal}{{\it Chem. Phys.\/}} \bibinfo{volume}{258}:\bibinfo{pages}{181}

\bibitem{RasaiahJC:Strast}
\bibinfo{author}{\bibnamefont{Rasaiah} \bibfnamefont{J}},
  \bibinfo{author}{\bibnamefont{Noworyta} \bibfnamefont{J}},
  \bibinfo{author}{\bibnamefont{Koneshan} \bibfnamefont{S.}} \bibinfo{year}{2000}.
  \bibinfo{journal}{{\it J. Am. Chem. Soc.\/}} \bibinfo{volume}{122}:\bibinfo{pages}{11182}

\bibitem{UrbicT:twomwT}
\bibinfo{author}{\bibnamefont{Urbic} \bibfnamefont{T}},
  \bibinfo{author}{\bibnamefont{Vlachy} \bibfnamefont{V}},
  \bibinfo{author}{\bibnamefont{Kalyuzhnyi} \bibfnamefont{Y}},
  \bibinfo{author}{\bibnamefont{Southall} \bibfnamefont{N}},
  \bibinfo{author}{\bibnamefont{Dill} \bibfnamefont{K.}}  \bibinfo{year}{2000}. 
  \bibinfo{journal}{{\it J.
  Chem. Phys.\/}} \bibinfo{volume}{112}: \bibinfo{pages}{2843}
  \bibinfo{year}{2000}.

\bibitem{NoworytaJP:Dynasi}
\bibinfo{author}{\bibnamefont{Noworyta} \bibfnamefont{J}},
  \bibinfo{author}{\bibnamefont{Koneshan} \bibfnamefont{S}},
  \bibinfo{author}{\bibnamefont{Rasaiah} \bibfnamefont{J.}} \bibinfo{year}{2000}.
  \bibinfo{journal}{{\it J. Am. Chem. Soc.\/}} \bibinfo{volume}{122}:\bibinfo{pages}{11194}

\bibitem{SchurhammerR:ArehA+}
\bibinfo{author}{\bibnamefont{Schurhammer} \bibfnamefont{R,}}
  \bibinfo{author}{\bibnamefont{Wipff} \bibfnamefont{G.}}  \bibinfo{year}{2000}. \bibinfo{journal}{{\it J.
  Phys. Chem. A\/}} \bibinfo{volume}{104}: \bibinfo{pages}{11159}


\bibitem{SouthallNT:mechsd}
\bibinfo{author}{\bibnamefont{Southall} \bibfnamefont{N,}}
  \bibinfo{author}{\bibnamefont{Dill} \bibfnamefont{K.}} \bibinfo{year}{2000}. \bibinfo{journal}{{\it J.
  Phys. Chem. B\/}} \bibinfo{volume}{104}: \bibinfo{pages}{1326}
  \bibinfo{year}{2000}.

\bibitem{GallicchioE:Entcda}
\bibinfo{author}{\bibnamefont{Gallicchio} \bibfnamefont{E}},
  \bibinfo{author}{\bibnamefont{Kubo} \bibfnamefont{MM}},
  \bibinfo{author}{\bibnamefont{Levy} \bibfnamefont{RM.}} \bibinfo{year}{2000}.
  \bibinfo{journal}{{\it J. Phys. Chem. B\/}} \bibinfo{volume}{104}:\bibinfo{pages}{6271}

\bibitem{Hernandez-CobosJ:hydhma}
\bibinfo{author}{\bibnamefont{Cobos} \bibfnamefont{J}},
  \bibinfo{author}{\bibnamefont{Mackie} \bibfnamefont{A}},
  \bibinfo{author}{\bibnamefont{Vega} \bibfnamefont{L.}} \bibinfo{year}{2001}. \bibinfo{journal}{{\it J.
  Chem. Phys.\/}} \bibinfo{volume}{114}: \bibinfo{pages}{7527}
 

\bibitem{BergmanDL:Isheuw}
\bibinfo{author}{\bibnamefont{Bergman} \bibfnamefont{D,}}
  \bibinfo{author}{\bibnamefont{Bell} \bibfnamefont{R.}} \bibinfo{year}{2001}.
  \bibinfo{journal}{{\it Mol. Phys.\/}}  \bibinfo{volume}{99}:\bibinfo{pages}{1011}

\bibitem{RaschkeTM:Quathi}
\bibinfo{author}{\bibnamefont{Raschke} \bibfnamefont{T}},
  \bibinfo{author}{\bibnamefont{Tsai} \bibfnamefont{J}},
  \bibinfo{author}{\bibnamefont{Levitt} \bibfnamefont{M.}} \bibinfo{year}{2001}.
  \bibinfo{journal}{{\it Proc. Nat. Acad. Sci. USA\/}} \bibinfo{volume}{98}:\bibinfo{pages}{5965}

\bibitem{GardeS:Temdhh}
\bibinfo{author}{\bibnamefont{Garde} \bibfnamefont{S,}}
  \bibinfo{author}{\bibnamefont{Ashbaugh} \bibfnamefont{HS.}} \bibinfo{year}{2001}.
  \bibinfo{journal}{{\it J. Chem. Phys.\/}} \bibinfo{volume}{115}:\bibinfo{pages}{977}

\bibitem{KairaA:Salshs}
\bibinfo{author}{\bibnamefont{Kaira} \bibfnamefont{A}},
  \bibinfo{author}{\bibnamefont{Tugcu} \bibfnamefont{N}},
  \bibinfo{author}{\bibnamefont{Cramer} \bibfnamefont{SM.}} \bibinfo{year}{2001}.
  \bibinfo{author}{\bibfnamefont{S.}~\bibnamefont{Garde}},
  \bibinfo{journal}{{\it J. Phys. Chem. B\/}} \bibinfo{volume}{105}:\bibinfo{pages}{6380}

\bibitem{Stillinger:73}
\bibinfo{author}{\bibnamefont{Stillinger} \bibfnamefont{FH.}} \bibinfo{year}{1973}.
  \bibinfo{journal}{{\it J. Sol. Chem.\/}} \bibinfo{volume}{2}:\bibinfo{pages}{141}


\bibitem{Pratt:91}
\bibinfo{author}{\bibnamefont{Pratt} \bibfnamefont{LR.}} \bibinfo{year}{1991}.  in
  \emph{\bibinfo{booktitle}{CLS Division 1991 Annual Review}}
  \bibinfo{publisher}{National Technical Information Service U. S. Department
  of Commerce}, \bibinfo{address}{5285 Port Royal Rd., Springfield, VA 22161},
  \bibinfo{note}{{L}A-UR-91-1783}.

\bibitem{TangKES:ExcvsS}
\bibinfo{author}{\bibnamefont{Tang} \bibfnamefont{K,}}
  \bibinfo{author}{\bibnamefont{Bloomfield} \bibfnamefont{V.}} \bibinfo{year}{2000}.
  \bibinfo{journal}{{\it Biophys. J.\/}} \bibinfo{volume}{79}:\bibinfo{pages}{2222}

\bibitem{Pierotti:76}
\bibinfo{author}{\bibnamefont{Pierotti} \bibfnamefont{RA.}} \bibinfo{year}{1976}.
  \bibinfo{journal}{{\it Chem. Rev.\/}} \bibinfo{volume}{76}:\bibinfo{pages}{717}

\bibitem{LUCASM:SIZETN}
\bibinfo{author}{\bibnamefont{Lucas} \bibfnamefont{M.}}  \bibinfo{year}{1976}. \bibinfo{journal}{{\it J.
  Phys. Chem.\/}} \bibinfo{volume}{80}: \bibinfo{pages}{359}


\bibitem{Ben-Naim:67}
\bibinfo{author}{\bibnamefont{Ben-Naim} \bibfnamefont{A,}}
  \bibinfo{author}{\bibnamefont{Friedman} \bibfnamefont{HL.}} \bibinfo{year}{1967}.
  \bibinfo{journal}{{\it J. Phys. Chem.\/}} \bibinfo{volume}{71}:\bibinfo{pages}{448}

\bibitem{Madan:BPC:94}
\bibinfo{author}{\bibnamefont{Madan} \bibfnamefont{B,}}
  \bibinfo{author}{\bibnamefont{Lee} \bibfnamefont{B.}} \bibinfo{year}{1994}.
  \bibinfo{journal}{{\it Biophys. Chem.\/}} \bibinfo{volume}{51}:\bibinfo{pages}{279}

\bibitem{Silverstein:JCE:98}
\bibinfo{author}{\bibnamefont{Silverstein} \bibfnamefont{TP.}} \bibinfo{year}{1998}.
  \bibinfo{journal}{{\it J. Chem. Ed.\/}} \bibinfo{volume}{75}:\bibinfo{pages}{116}

\bibitem{Lazaridis:2001}
\bibinfo{author}{\bibnamefont{Lazaridis} \bibfnamefont{T.}} \bibinfo{year}{2001}.
  \emph{\bibinfo{title}{Solvent size vs cohesive energy as the origin of
  hydrophobicity}}, \bibinfo{note}{private communication}

\bibitem{Rowlinson:Swinton:82}
\bibinfo{author}{\bibnamefont{Rowlinson} \bibfnamefont{JS,}}
  \bibinfo{author}{\bibnamefont{Swinton},
  \bibfnamefont{FL.}} \bibinfo{year}{1982}. \emph{\bibinfo{title}{Liquids and Liquid
  Mixtures}}, \bibinfo{publisher}{Butterworths}, \bibinfo{address}{London}
   


\bibitem{Postma:FSCS:82}
\bibinfo{author}{\bibnamefont{Postma} \bibfnamefont{JPM}},
  \bibinfo{author}{\bibnamefont{Berendsen} \bibfnamefont{HJC}},
  \bibinfo{author}{\bibnamefont{Haak} \bibfnamefont{JR.}} \bibinfo{year}{1982}.
  \bibinfo{journal}{{\it Faraday Symp. Chem. Soc.\/}} \bibinfo{pages}{55}
 

\bibitem{POHORILLEA:MOLAI}
\bibinfo{author}{\bibnamefont{Pohorille} \bibfnamefont{A,}}
  \bibinfo{author}{\bibnamefont{Wilson} \bibfnamefont{M.}} \bibinfo{year}{1993}.
  \bibinfo{journal}{{\it J. Mol. Struct. (Theochem)\/}} \bibinfo{volume}{103}:\bibinfo{pages}{271}

\bibitem{Wolfenden:94}
\bibinfo{author}{\bibnamefont{Wolfenden} \bibfnamefont{R,}}
  \bibinfo{author}{\bibnamefont{Radzicka} \bibfnamefont{A.}~} \bibinfo{year}{1994}.
  \bibinfo{journal}{{\it Science\/}} \bibinfo{volume}{265}:\bibinfo{pages}{936}

\bibitem{Kocher:Structure:96}
\bibinfo{author}{\bibnamefont{Kocher} \bibfnamefont{JP}},
  \bibinfo{author}{\bibnamefont{Prevost} \bibfnamefont{M}},
  \bibinfo{author}{\bibnamefont{Wodak} \bibfnamefont{SJ}},
  \bibinfo{author}{\bibnamefont{Lee} \bibfnamefont{B.}} \bibinfo{year}{1996}.
  \bibinfo{journal}{{\it Structure}} \bibinfo{volume}{4}:\bibinfo{pages}{1517}

\bibitem{Crooks:PRE:97}
\bibinfo{author}{\bibnamefont{Crooks} \bibfnamefont{GE,}}
  \bibinfo{author}{\bibnamefont{Chandler} \bibfnamefont{D.}} \bibinfo{year}{1997}.
  \bibinfo{journal}{{\it Phys. Rev. E\/}} \bibinfo{volume}{56}:\bibinfo{pages}{4217}

\bibitem{Stamatopoulou:JCP:98}
\bibinfo{author}{\bibnamefont{Stamatopoulou} \bibfnamefont{A,}}
  \bibinfo{author}{\bibnamefont{Ben-Amotz} \bibfnamefont{D.}~} \bibinfo{year}{1998}.
  \bibinfo{journal}{{\it J. Chem. Phys.\/}} \bibinfo{volume}{108}:\bibinfo{pages}{7294}

\bibitem{TanakaH:Cavdlw}
\bibinfo{author}{\bibnamefont{Tanaka} \bibfnamefont{H.}} \bibinfo{year}{1998}.
  \bibinfo{journal}{{\it Chem. Phys. Letts.\/}} \bibinfo{volume}{282}:\bibinfo{pages}{133}

\bibitem{MountainRD:Voicew}
\bibinfo{author}{\bibnamefont{Mountain} \bibfnamefont{R.}} \bibinfo{year}{1999}.
  \bibinfo{journal}{{\it J. Chem. Phys.\/}} \bibinfo{volume}{110}:\bibinfo{pages}{2109}

\bibitem{GardeS:Micdfs}
\bibinfo{author}{\bibnamefont{Garde} \bibfnamefont{S}},
  \bibinfo{author}{\bibnamefont{Khare} \bibfnamefont{R}},
  \bibinfo{author}{\bibnamefont{Hummer} \bibfnamefont{G.}} \bibinfo{year}{2000}.
  \bibinfo{journal}{{\it J. Chem. Phys.\/}} \bibinfo{volume}{112}:\bibinfo{pages}{1574}

\bibitem{intVeldPJI:Liqsvc}
\bibinfo{author}{\bibnamefont{in'tVeld} \bibfnamefont{PJ}},
  \bibinfo{author}{\bibnamefont{Stone} \bibfnamefont{M}},
  \bibinfo{author}{\bibnamefont{Truskett} \bibfnamefont{T}},
  \bibinfo{author}{\bibnamefont{Sanchez}} \bibfnamefont{I.} \bibinfo{year}{2000}.
  \bibinfo{journal}{{\it J. Phys. Chem. B\/}} \bibinfo{volume}{104}:\bibinfo{pages}{12028}

\bibitem{KussellE:Excvps}
\bibinfo{author}{\bibnamefont{Kussell} \bibfnamefont{E}},
  \bibinfo{author}{\bibnamefont{Shimada} \bibfnamefont{J}},
  \bibinfo{author}{\bibnamefont{Shakhnovich} \bibfnamefont{E.}} \bibinfo{year}{2001}.
  \bibinfo{journal}{{\it J. Mol. Bio.\/}} \bibinfo{volume}{311}:\bibinfo{pages}{183}
  
\bibitem{LesemannM:Predtc}
\bibinfo{author}{\bibnamefont{Lesemann} \bibfnamefont{M}},
\bibinfo{author}{\bibnamefont{Thirumoorthy} \bibfnamefont{K}},
\bibinfo{author}{\bibnamefont{Kim} \bibfnamefont{YJ}},
\bibinfo{author}{\bibnamefont{Jonas} \bibfnamefont{J}},
\bibinfo{author}{\bibnamefont{Paulaitis} \bibfnamefont{ME}}.
\bibinfo{year}{1998}. \bibinfo{journal}{{\it Langmuir\/}} \bibinfo{volume}{14}:
\bibinfo{pages}{5339--5341}

\bibitem{Percus:JP:93}
\bibinfo{author}{\bibnamefont{Percus} \bibfnamefont{JK.}} \bibinfo{year}{1993}.
  \bibinfo{journal}{{\it J. Physique IV\/}} \bibinfo{volume}{3}:\bibinfo{pages}{49}

\bibitem{MITCHELLDJ:HARSEC}
\bibinfo{author}{\bibnamefont{Mitchell} \bibfnamefont{D}},
  \bibinfo{author}{\bibnamefont{Ninham} \bibfnamefont{B}},
  \bibinfo{author}{\bibnamefont{Pailthorpe} \bibfnamefont{B.}} \bibinfo{year}{1977}.
  \bibinfo{journal}{{\it Chem. Phys. Letts.\/}} \bibinfo{volume}{51}:\bibinfo{pages}{257}

\bibitem{Stell:85}
\bibinfo{author}{\bibnamefont{Stell} \bibfnamefont{G.}} \bibinfo{year}{1985}.  in
  \emph{\bibinfo{booktitle}{The Wonderful World of Stochastics. A Tribute to
  Elliot W. Montroll}}, edited by \bibinfo{editor}{\bibfnamefont{MF}
  \bibnamefont{Schlesinger}} and 
  \bibinfo{editor}{\bibfnamefont{GH} \bibnamefont{Weiss}}
  \bibinfo{publisher}{Elsevier}, \bibinfo{address}{NY},
  \bibinfo{volume}{XII}: \bibinfo{pages}{127}, \bibinfo{note}{{S}tudies
  in Statistical Mechanics}.



\bibitem{Widom:JCP:63}
\bibinfo{author}{\bibnamefont{Widom} \bibfnamefont{B.}} \bibinfo{year}{1963}. \bibinfo{journal}{{\it J.
  Chem. Phys.\/}} \bibinfo{volume}{39}: \bibinfo{pages}{2808}
  \bibinfo{year}{1963}.

\bibitem{Torrie:JCompP:77}
\bibinfo{author}{\bibnamefont{Torrie} \bibfnamefont{GM.}}
  \bibinfo{author}{\bibnamefont{Valleau} \bibfnamefont{JP.}} \bibinfo{year}{1977}.
  \bibinfo{journal}{{\it J. Comp. Phys.\/}} \bibinfo{volume}{23}:\bibinfo{pages}{187}

\bibitem{Valleau}
\bibinfo{author}{\bibnamefont{Valleau} \bibfnamefont{JP.}}
  \bibinfo{author}{\bibnamefont{Torrie} \bibfnamefont{GM.}} \bibinfo{year}{1977}. in
  \emph{\bibinfo{booktitle}{Statistical Mechanics, Part A: Equilibrium
  Techniques}}, edited by \bibinfo{editor}{\bibfnamefont{BJ.}
  \bibnamefont{Berne}}, \bibinfo{publisher}{Plenum}, \bibinfo{address}{New
  York},  \bibinfo{volume}{5}:  \bibinfo{pages}{169}.
  
\bibitem{weeks:ARPC2001}
\bibinfo{author}{\bibnamefont{Weeks} \bibfnamefont{JD.}} \bibinfo{year}{2002}. 
  \bibinfo{journal}{{\it Ann. Rev. Phys. Chem\/}} \bibinfo{volume}{52}:\bibinfo{pages}{xxx}

  
  \bibitem{Ashbaugh:2001}
\bibinfo{author}{\bibnamefont{Ashbuagh} \bibfnamefont{HS,}}
  \bibinfo{author}{\bibnamefont{Paulaitis} \bibfnamefont{ME.}} \bibinfo{year}{in press 2001}.
  \bibinfo{journal}{{\it J. Am. Chem. Soc.\/}} \bibinfo{volume}{xxx}:\bibinfo{pages}{yyy}

  
  \bibitem{Chandler:83}
\bibinfo{author}{\bibnamefont{Chandler} \bibfnamefont{D}},
  \bibinfo{author}{\bibnamefont{Weeks} \bibfnamefont{JD}},
  \bibinfo{author}{\bibnamefont{Andersen} \bibfnamefont{HC.}} \bibinfo{year}{1983}.
  \bibinfo{journal}{{\it Science\/}} \bibinfo{volume}{220}:\bibinfo{pages}{787}

  
  \bibitem{Stell:77}
\bibinfo{author}{\bibnamefont{Stell} \bibfnamefont{G.}} \bibinfo{year}{1977}.  in
  \emph{\bibinfo{booktitle}{Statistical Mechanics, Part A: Equilibrium
  Techniques}}, edited by \bibinfo{editor}{\bibfnamefont{BJ.}
  \bibnamefont{Berne}}, \bibinfo{publisher}{Plenum}, \bibinfo{address}{NY},
   \bibinfo{volume}{5}:  \bibinfo{pages}{47},
  \bibinfo{note}{{M}odern Theoretical Chemistry}.


\bibitem{POLLACKGL:WHYGDL}
\bibinfo{author}{\bibnamefont{Pollack} \bibfnamefont{G.}} \bibinfo{year}{1991}.
  \bibinfo{journal}{{\it Science\/}} \bibinfo{volume}{251}:\bibinfo{pages}{1323}
  
\bibitem{Baldwin:2001}
\bibinfo{author}{\bibnamefont{Baldwin} \bibfnamefont{RL.}} \bibinfo{year}{1986}.
\bibinfo{journal}{{\it Proc.
  Nat. Acad. Sci. USA\/}} \bibinfo{volume}{83}: \bibinfo{pages}{8069}

\bibitem{LEEB:ISOITT}
\bibinfo{author}{\bibnamefont{Lee} \bibfnamefont{B.}}  \bibinfo{year}{1991}.
 \bibinfo{journal}{{\it Proc.
  Nat. Acad. Sci. USA\/}} \bibinfo{volume}{88}: \bibinfo{pages}{5154}
  
  

\bibitem{BoulougourisGC:Hencaw}
\bibinfo{author}{\bibnamefont{Boulougouris} \bibfnamefont{GC,}}
\bibinfo{author}{\bibnamefont{Voutsas} \bibfnamefont{EC,}}
\bibinfo{author}{\bibnamefont{Economou} \bibfnamefont{IG,}}
\bibinfo{author}{\bibnamefont{Theodorou} \bibfnamefont{DN,}}
\bibinfo{author}{\bibnamefont{Tassios} \bibfnamefont{DP.}} \bibinfo{year}{2001}
 \bibinfo{journal}{{\it J. Phys. Chem. B\/}} 
 \bibinfo{volume}{105}:
\bibinfo{pages}{7792--7798}

\bibitem{Ashbaugh:2001b}
\bibinfo{author}{\bibnamefont{Ashbaugh}} \bibfnamefont{HS},
  \bibinfo{author}{\bibnamefont{Truskett} \bibfnamefont{TM}},
  \bibinfo{author}{\bibnamefont{Debenedetti} \bibfnamefont{PG.}}  \bibinfo{year}{2001}.
\emph{\bibinfo{title}{A simple molecular thermodynamic theory of hydrophobic hydration}},  
  \bibinfo{journal}{ private communication} 


\bibitem{Filipponi:97}
\bibinfo{author}{\bibnamefont{Filipponi} \bibfnamefont{A}},
  \bibinfo{author}{\bibnamefont{Bowron} \bibfnamefont{DT}},
  \bibinfo{author}{\bibnamefont{Lobban} \bibfnamefont{C}},
  \bibinfo{author}{\bibnamefont{Finney} \bibfnamefont{JL.}} \bibinfo{year}{1997}.
  \bibinfo{journal}{{\it Phys. Rev. Lett.\/}} \bibinfo{volume}{79}:\bibinfo{pages}{1293}

\bibitem{BowronDT:Temdth}
\bibinfo{author}{\bibnamefont{Bowron} \bibfnamefont{D}},
  \bibinfo{author}{\bibnamefont{Filipponi} \bibfnamefont{A}},
  \bibinfo{author}{\bibnamefont{Lobban} \bibfnamefont{C}},
  \bibinfo{author}{\bibnamefont{Finney} \bibfnamefont{J.}} \bibinfo{year}{1998}.
  \bibinfo{journal}{{\it Chem. Phys. Letts.\/}} \bibinfo{volume}{293}:\bibinfo{pages}{33} 

\bibitem{BowronDT:Hydhfc}
\bibinfo{author}{\bibnamefont{Bowron} \bibfnamefont{D}},
  \bibinfo{author}{\bibnamefont{Filipponi} \bibfnamefont{A}},
  \bibinfo{author}{\bibnamefont{Roberts} \bibfnamefont{M}},
  \bibinfo{author}{\bibnamefont{Finney} \bibfnamefont{J.}} \bibinfo{year}{1998}.
  \bibinfo{journal}{{\it Phys. Rev. Letts.\/}} \bibinfo{volume}{81}:\bibinfo{pages}{4164}

\bibitem{BowronDT:Xraasi}
\bibinfo{author}{\bibnamefont{Bowron} \bibfnamefont{D}},
  \bibinfo{author}{\bibnamefont{Weigel} \bibfnamefont{R}},
  \bibinfo{author}{\bibnamefont{Filipponi} \bibfnamefont{A}},
  \bibinfo{author}{\bibnamefont{Roberts} \bibfnamefont{M}},
  \bibinfo{author}{\bibnamefont{Finney} \bibfnamefont{J.}} \bibinfo{year}{2001}.
  \bibinfo{journal}{{\it Mol. Phys.\/}}  \bibinfo{volume}{99}:\bibinfo{pages}{761}
  
\bibitem{SullivanDM:Hydhah}
\bibinfo{author}{\bibnamefont{Sullivan} \bibfnamefont{D}},
  \bibinfo{author}{\bibnamefont{Neilson} \bibfnamefont{G}},
  \bibinfo{author}{\bibnamefont{Fischer} \bibfnamefont{H.}} \bibinfo{year}{2001}.
  \bibinfo{journal}{{\it J. Chem. Phys.\/}} \bibinfo{volume}{115}:\bibinfo{pages}{339}

\bibitem{GHOSHT:2001a}
\bibinfo{author}{\bibnamefont{Ghosh} \bibfnamefont{T}},
  \bibinfo{author}{\bibnamefont{Garde} \bibfnamefont{S}},
  \bibinfo{author}{\bibnamefont{Garc\'{i}a} \bibfnamefont{AE.}} \bibinfo{year}{in press 2001}.
  \bibinfo{journal}{{\it J. Am. Chem. Soc.\/}} \bibinfo{volume}{xxx}:\bibinfo{pages}{yyy}

\bibitem{Frank:JCP:45}
\bibinfo{author}{\bibnamefont{Frank} \bibfnamefont{HS,}}
  \bibinfo{author}{\bibnamefont{Evans} \bibfnamefont{MW.}} \bibinfo{year}{1945}.
  \bibinfo{journal}{{\it J. Chem. Phys.\/}} \bibinfo{volume}{13}:\bibinfo{pages}{507}

\bibitem{pellegrinim:modscc}
\bibinfo{author}{\bibnamefont{Pellegrini} \bibfnamefont{M,}}
  \bibinfo{author}{\bibnamefont{Doniach} \bibfnamefont{S.}} \bibinfo{year}{1995}.
  \bibinfo{journal}{{\it J. Chem. Phys.\/}} \bibinfo{volume}{103}:\bibinfo{pages}{2696}

\bibitem{pellegrinim:potmfb}
\bibinfo{author}{\bibnamefont{Pellegrini} \bibfnamefont{M}},
  \bibinfo{author}{\bibnamefont{Gronbechjensen} \bibfnamefont{N}},
  \bibinfo{author}{\bibnamefont{Doniach} \bibfnamefont{S.}} \bibinfo{year}{1996}.
  \bibinfo{journal}{{\it J. Chem. Phys.\/}} \bibinfo{volume}{104}:\bibinfo{pages}{8639}

\bibitem{AshbaughHS:Coneaa}
\bibinfo{author}{\bibnamefont{Ashbaugh} \bibfnamefont{HS}},
  \bibinfo{author}{\bibnamefont{Garde} \bibfnamefont{S}},
  \bibinfo{author}{\bibnamefont{Hummer} \bibfnamefont{G}},
  \bibinfo{author}{\bibnamefont{Kaler} \bibfnamefont{EW}},
  \bibinfo{author}{\bibnamefont{Paulaitis} \bibfnamefont{ME.}} \bibinfo{year}{1999}.
  \bibinfo{journal}{{\it Biophys. J.\/}} \bibinfo{volume}{77}:\bibinfo{pages}{645} \bibinfo{year}{1999}.

\bibitem{wallqvista:molsha}
\bibinfo{author}{\bibnamefont{Wallqvist} \bibfnamefont{A.}} \bibinfo{year}{1991}.
  \bibinfo{journal}{{\it J. Phys. Chem.\/}} \bibinfo{volume}{95}:\bibinfo{pages}{8921}

\bibitem{ChengYK:Surtdb}
\bibinfo{author}{\bibnamefont{Cheng} \bibfnamefont{Y,}}
  \bibinfo{author}{\bibnamefont{Rossky} \bibfnamefont{P.}} \bibinfo{year}{1998}.
  \bibinfo{journal}{{\it Nature\/}} \bibinfo{volume}{392}:\bibinfo{pages}{696}

\bibitem{PANGALICS:DETTM2}
\bibinfo{author}{\bibnamefont{Pangali} \bibfnamefont{C}},
  \bibinfo{author}{\bibnamefont{Rao} \bibfnamefont{M}},
  \bibinfo{author}{\bibnamefont{Berne} \bibfnamefont{B.}} \bibinfo{year}{1978}.
  \bibinfo{journal}{{\it {ACS} Symposium Series\/}} \bibinfo{volume}{1978}:\bibinfo{pages}{32} 

\bibitem{Pangali:79}
\bibinfo{author}{\bibnamefont{Pangali} \bibfnamefont{C}},
  \bibinfo{author}{\bibnamefont{Rao} \bibfnamefont{M}},
  \bibinfo{author}{\bibnamefont{Berne} \bibfnamefont{BJ.}} \bibinfo{year}{1979}.
  \bibinfo{journal}{{\it J. Chem. Phys.\/}} \bibinfo{volume}{71}:\bibinfo{pages}{2975}

\bibitem{swaminathans:monchb}
\bibinfo{author}{\bibnamefont{Swaminathan} \bibfnamefont{S,}}
  \bibinfo{author}{\bibnamefont{Beveridge} \bibfnamefont{DL.}} \bibinfo{year}{1979}.
  \bibinfo{journal}{{\it J. Am. Chem. Soc.\/}} \bibinfo{volume}{101}:\bibinfo{pages}{5832}

\bibitem{ravishankerg:moncst}
\bibinfo{author}{\bibnamefont{Ravishanker} \bibfnamefont{G}},
  \bibinfo{author}{\bibnamefont{Mezei} \bibfnamefont{M}},
  \bibinfo{author}{\bibnamefont{Beveridge} \bibfnamefont{DL.}}  \bibinfo{year}{1982}.
  \bibinfo{journal}{{\it Faraday Symp. Chem. Soc.\/}}  \bibinfo{pages}{79}


\bibitem{BEVERIDGEDL:FRESAT}
\bibinfo{author}{\bibnamefont{Beveridge} \bibfnamefont{DL}},
  \bibinfo{author}{\bibnamefont{Mezei} \bibfnamefont{M}},
  \bibinfo{author}{\bibnamefont{Ravishanker} \bibfnamefont{G.}} \bibinfo{year}{1985}.
  \bibinfo{author}{\bibnamefont{Jayaram} \bibfnamefont{B}},
  \bibinfo{journal}{{\it J. Biosci.\/}} \bibinfo{volume}{8}:\bibinfo{pages}{167}

\bibitem{RAVISHANKERG:POTMFS}
\bibinfo{author}{\bibnamefont{Ravishanker} \bibfnamefont{G,}}
  \bibinfo{author}{\bibnamefont{Beveridge} \bibfnamefont{DL.}} \bibinfo{year}{1985}.
  \bibinfo{journal}{{\it J. Am. Chem. Soc.\/}} \bibinfo{volume}{107}:\bibinfo{pages}{2565}

\bibitem{backxp:signrf}
\bibinfo{author}{\bibnamefont{Backx} \bibfnamefont{P,}}
  \bibinfo{author}{\bibnamefont{Goldman} \bibfnamefont{S.}} \bibinfo{year}{1985}.
  \bibinfo{journal}{{\it Chem. Phys. Letts.\/}} \bibinfo{volume}{113}:\bibinfo{pages}{578}

\bibitem{backxp:somcrr}
\bibinfo{author}{\bibnamefont{Backx} \bibfnamefont{P,}}
  \bibinfo{author}{\bibnamefont{Goldman} \bibfnamefont{S.}} \bibinfo{year}{1985}.
  \bibinfo{journal}{{\it Chem. Phys. Letts.\/}} \bibinfo{volume}{119}:\bibinfo{pages}{144} \bibinfo{year}{1985}.

\bibitem{WATANABEK:MOLSTH}
\bibinfo{author}{\bibnamefont{Watanabe} \bibfnamefont{K,}}
  \bibinfo{author}{\bibnamefont{Andersen} \bibfnamefont{HC.}} \bibinfo{year}{1986}.
  \bibinfo{journal}{{\it J. Phys. Chem.\/}} \bibinfo{volume}{90}:\bibinfo{pages}{795}

\bibitem{watanabe:86}
\bibinfo{author}{\bibnamefont{Watanabe} \bibfnamefont{K,}}
  \bibinfo{author}{\bibnamefont{Andersen} \bibfnamefont{HC.}} \bibinfo{year}{1986}. in
  \emph{\bibinfo{booktitle}{Molecular-Dynamics Simulation of
  Statistical-Mechanical Systems}}, edited by
  \bibinfo{editor}{\bibfnamefont{G.}~\bibnamefont{Ciccotti}}
  \bibinfo{editor}{\bibfnamefont{WG.} \bibnamefont{Hoover}}
  \emph{\bibinfo{series}{ASI-NATO International School of Physics ``Enrico
  Fermi''}}, \bibinfo{volume}{XCVII}: \bibinfo{pages}{418}.

\bibitem{WALLQVISTA:HYDIBM}
\bibinfo{author}{\bibnamefont{Wallqvist} \bibfnamefont{A,}}
  \bibinfo{author}{\bibnamefont{Berne} \bibfnamefont{B.}} \bibinfo{year}{1988}.
  \bibinfo{journal}{{\it Chem. Phys. Letts.\/}} \bibinfo{volume}{145}:\bibinfo{pages}{26}

\bibitem{JORGENSENWL:EFFCAF}
\bibinfo{author}{\bibnamefont{Jorgensen} \bibfnamefont{WL}},
  \bibinfo{author}{\bibnamefont{Buckner} \bibfnamefont{JK}},
  \bibinfo{author}{\bibnamefont{Boudon} \bibfnamefont{S}},
  \bibinfo{author}{\bibnamefont{Tiradorives} \bibfnamefont{J.}} \bibinfo{year}{1988}.
  \bibinfo{journal}{{\it J. Chem. Phys.\/}} \bibinfo{volume}{89}:\bibinfo{pages}{3742}

\bibitem{BERNEBJ:MODHI}
\bibinfo{author}{\bibnamefont{Berne} \bibfnamefont{B,}}
  \bibinfo{author}{\bibnamefont{Wallqvist} \bibfnamefont{A.}} \bibinfo{year}{1989}.
  \bibinfo{journal}{{\it Chem. Scr.\/}} \bibinfo{volume}{29A}:\bibinfo{pages}{85}
  
  
\bibitem{Rossky:80}
\bibinfo{author}{ \bibnamefont{Rossky} \bibfnamefont{PJ}},
  \bibinfo{author}{ \bibnamefont{Friedman} \bibfnamefont{HL.}} \bibinfo{year}{1980}.
  \bibinfo{journal}{{\it J. Phys. Chem.\/}} \bibinfo{volume}{84}:\bibinfo{pages}{587}

  
\bibitem{JORGENSENWL:AAPMF}
\bibinfo{author}{\bibnamefont{Jorgensen} \bibfnamefont{WL,}}
\bibinfo{author}{\bibnamefont{Severance} \bibfnamefont{DL.}}
\bibinfo{year}{1990}.
  \bibinfo{journal}{{\it J. Am. Chem. Soc.\/}} \bibinfo{volume}{112}:\bibinfo{pages}{4786}
 


\bibitem{LINSEP:STATBD}
\bibinfo{author}{\bibnamefont{Linse} \bibfnamefont{P.}} \bibinfo{year}{1992}. \bibinfo{journal}{{\it J.
  Am. Chem. Soc.\/}} \bibinfo{volume}{114}: \bibinfo{pages}{4366}
 
\bibitem{LINSEP:ORIBBP}
\bibinfo{author}{\bibnamefont{Linse} \bibfnamefont{P.}}  \bibinfo{year}{1993}. \bibinfo{journal}{{\it J.
  Am. Chem. Soc.\/}} \bibinfo{volume}{115}: \bibinfo{pages}{8793}




\bibitem{Smith:JACS:92}
\bibinfo{author}{\bibnamefont{Smith} \bibfnamefont{DE}},
  \bibinfo{author}{\bibnamefont{Zhang} \bibfnamefont{L}},
  \bibinfo{author}{\bibnamefont{Haymet} \bibfnamefont{ADJ.}} \bibinfo{year}{1992}.
  \bibinfo{journal}{{\it J. Am. Chem. Soc.\/}} \bibinfo{volume}{114}:\bibinfo{pages}{5875}

\bibitem{Smith:JCP:93}
\bibinfo{author}{\bibnamefont{Smith} \bibfnamefont{DE,}}
  \bibinfo{author}{\bibnamefont{Haymet} \bibfnamefont{ADJ.}} \bibinfo{year}{1993}.
  \bibinfo{journal}{{\it J. Chem. Phys.\/}} \bibinfo{volume}{98}:\bibinfo{pages}{6445}
  
\bibitem{danglx:potmfm}
\bibinfo{author}{\bibnamefont{Dang} \bibfnamefont{LX.}}  \bibinfo{year}{1994}. \bibinfo{journal}{{\it J.
  Chem. Phys.\/}} \bibinfo{volume}{100}: \bibinfo{pages}{9032}


\bibitem{newmh:molcte}
\bibinfo{author}{\bibnamefont{New} \bibfnamefont{MH,}}
  \bibinfo{author}{\bibnamefont{Berne} \bibfnamefont{BJ.}} \bibinfo{year}{1995}.
  \bibinfo{journal}{{\it J. Am. Chem. Soc.\/}} \bibinfo{volume}{117}:\bibinfo{pages}{7172}

\bibitem{LudemannS:inftph}
\bibinfo{author}{\bibnamefont{{L\"{u}demann}} \bibfnamefont{S}},
  \bibinfo{author}{\bibnamefont{Schreiber} \bibfnamefont{H}},
  \bibinfo{author}{\bibnamefont{Abseher} \bibfnamefont{R}},
  \bibinfo{author}{\bibnamefont{Steinhauser} \bibfnamefont{O.}} \bibinfo{year}{1996}.
  \bibinfo{journal}{{\it J. Chem. Phys.\/}} \bibinfo{volume}{104}:\bibinfo{pages}{286}

\bibitem{Ludemann:97}
\bibinfo{author}{\bibnamefont{{L\"{u}demann} \bibfnamefont{S}}},
  \bibinfo{author}{\bibnamefont{Abseher} \bibfnamefont{R}},
  \bibinfo{author}{\bibnamefont{Schreiber} \bibfnamefont{H}},
  \bibinfo{author}{\bibnamefont{Steinhauser} \bibfnamefont{O.}} \bibinfo{year}{1997}.
  \bibinfo{journal}{{\it J. Am. Chem. Soc.\/}} \bibinfo{volume}{119}:\bibinfo{pages}{4206}

\bibitem{Payne:97}
\bibinfo{author}{\bibnamefont{Payne} \bibfnamefont{VA}},
  \bibinfo{author}{\bibnamefont{Matubayasi} \bibfnamefont{N}},
  \bibinfo{author}{\bibnamefont{Murphy} \bibfnamefont{LR}},
  \bibinfo{author}{\bibnamefont{Levy} \bibfnamefont{RM.}} \bibinfo{year}{1997}.
  \bibinfo{journal}{{\it J. Phys. Chem. B\/}} \bibinfo{volume}{101}:\bibinfo{pages}{2054}

\bibitem{youngws:reethe}
\bibinfo{author}{\bibnamefont{Young} \bibfnamefont{WS,}}
  \bibinfo{author}{\bibnamefont{Brooks} \bibfnamefont{CL.}} \bibinfo{year}{1997}.
  \bibinfo{journal}{{\it J. Chem. Phys.\/}} \bibinfo{volume}{106}:\bibinfo{pages}{9265}

\bibitem{Rick:97}
\bibinfo{author}{\bibnamefont{Rick} \bibfnamefont{SW,}}
  \bibinfo{author}{\bibnamefont{Rick} \bibfnamefont{BJ.}} \bibinfo{year}{1997}.
  \bibinfo{journal}{{\it J. Phys. Chem. B\/}} \bibinfo{volume}{101}:\bibinfo{pages}{10488}

\bibitem{RankJA:Conshb}
\bibinfo{author}{\bibnamefont{Rank} \bibfnamefont{JA.}}
  \bibinfo{author}{\bibnamefont{Baker} \bibfnamefont{D.}} \bibinfo{year}{1998}.
  \bibinfo{journal}{{\it Biophys. Chem.\/}} \bibinfo{volume}{71}:\bibinfo{pages}{199}
  
\bibitem{GAOJL:SUPHOP}
\bibinfo{author}{\bibnamefont{Gao} \bibfnamefont{JL}}.  \bibinfo{year}{1993}.
  \bibinfo{journal}{{\it J. Am. Chem. Soc.\/}} \bibinfo{volume}{115}:\bibinfo{pages}{6893--6895},

\bibitem{ChipotC:Bendgm}
\bibinfo{author}{\bibnamefont{Chipot} \bibfnamefont{C}},
  \bibinfo{author}{\bibnamefont{Jaffe} \bibfnamefont{R}},
  \bibinfo{author}{\bibnamefont{Maigret} \bibfnamefont{B}},
  \bibinfo{author}{\bibnamefont{Pearlman} \bibfnamefont{D}},
  \bibinfo{author}{\bibnamefont{Kollman} \bibfnamefont{P.}} \bibinfo{year}{1996}.
  \bibinfo{journal}{{\it J. Am. Chem. Soc.\/}} \bibinfo{volume}{118}:\bibinfo{pages}{11217}.

\bibitem{Rick:2000}
\bibinfo{author}{\bibnamefont{Rick} \bibfnamefont{SW,}}
\bibinfo{year}{2000}. \bibinfo{journal}{{\it J. Phys. Chem. B\/}}
\bibinfo{volume}{104}: \bibinfo{pages}{6884}


\bibitem{ShimizuS:Temdhi}
\bibinfo{author}{\bibnamefont{Shimizu} \bibfnamefont{S,}}
  \bibinfo{author}{\bibnamefont{Chan} \bibfnamefont{H.}} \bibinfo{year}{2000}. \bibinfo{journal}{{\it J.
  Chem. Phys.\/}} \bibinfo{volume}{113}: \bibinfo{pages}{4683}
  
 \bibitem{Gervazi:AAPMF}
\bibinfo{author}{\bibnamefont{Gervazi} \bibfnamefont{FL}},
  \bibinfo{author}{\bibnamefont{Chelli} \bibfnamefont{R}},
  \bibinfo{author}{\bibnamefont{Marchi} \bibfnamefont{M}},
  \bibinfo{author}{\bibnamefont{Procacci} \bibfnamefont{P}},
  \bibinfo{author}{\bibnamefont{Scettino} \bibfnamefont{V}} \bibinfo{year}{2001}.
  \bibinfo{journal}{{\it J. Phys. Chem. B\/}} \bibinfo{volume}{105}:\bibinfo{pages}{7835}.

 

\bibitem{GHOSHT:2001b}
\bibinfo{author}{\bibnamefont{Ghosh} \bibfnamefont{T}},
  \bibinfo{author}{\bibnamefont{Garde} \bibfnamefont{S}},
  \bibinfo{author}{\bibnamefont{Garc\'{i}a} \bibfnamefont{AE.}} \bibinfo{year}{2001 (in press)}.
  \bibinfo{journal}{{\it J. Chem. Phys.\/}} \bibinfo{volume}{xxx}:\bibinfo{pages}{yyy} 

\bibitem{Pratt:80b}
\bibinfo{author}{\bibnamefont{Pratt} \bibfnamefont{LR,}}
  \bibinfo{author}{\bibnamefont{Chandler} \bibfnamefont{D.}} \bibinfo{year}{1980}.
  \bibinfo{journal}{{\it J. Chem. Phys.\/}} \bibinfo{volume}{73}:\bibinfo{pages}{3434}

\bibitem{Pratt:85a}
\bibinfo{author}{\bibnamefont{Pratt} \bibfnamefont{LR.}} \bibinfo{year}{1985}.
  \bibinfo{journal}{{\it Ann. Rev. Phys. Chem.\/}} \bibinfo{volume}{36}:\bibinfo{pages}{433}

\bibitem{HummerG:FastiE} \bibinfo{author}{\bibnamefont{Hummer}
\bibfnamefont{G.}}  \bibinfo{year}{2001}. \bibinfo{journal}{{\it J.
Chem. Phys.\/}} \bibinfo{volume}{114}: \bibinfo{pages}{7330}


\bibitem{jorgensen_1982}
\bibinfo{author}{\bibnamefont{Jorgensen} \bibfnamefont{WL.}} \bibinfo{year}{1982}.
  \bibinfo{journal}{{\it J. Chem. Phys.\/}} \bibinfo{volume}{77}:\bibinfo{pages}{5757}

\bibitem{berne7_1982}
\bibinfo{author}{\bibnamefont{Rosenberg} \bibfnamefont{RO}},
  \bibinfo{author}{\bibnamefont{Mikkilineni} \bibfnamefont{R}},
  \bibinfo{author}{\bibnamefont{Berne} \bibfnamefont{BJ.}} \bibinfo{year}{1982}.
  \bibinfo{journal}{{\it J. Am. Chem. Soc.\/}} \bibinfo{volume}{104}:\bibinfo{pages}{7647}

\bibitem{Buckner:87}
\bibinfo{author}{\bibnamefont{Jorgensen} \bibfnamefont{WL,}}
  \bibinfo{author}{\bibnamefont{Buckner} \bibinfo{year}{1987}. \bibfnamefont{JK}}, 
  \bibinfo{journal}{{\it J. Phys. Chem.\/}}
  \bibinfo{volume}{91}: \bibinfo{pages}{6083}

\bibitem{Tobias:90}
\bibinfo{author}{\bibnamefont{Tobias} \bibfnamefont{DJ,}}
  \bibinfo{author}{\bibnamefont{Brooks} \bibfnamefont{CL.}} \bibinfo{year}{1990}.
  \bibinfo{journal}{{\it J. Chem. Phys.\/}} \bibinfo{volume}{92}:\bibinfo{pages}{2582}

\bibitem{Wallqvist:95}
\bibinfo{author}{\bibnamefont{Wallqvist} \bibfnamefont{A,}}
  \bibinfo{author}{\bibnamefont{Covell} \bibfnamefont{DG.}} \bibinfo{year}{1995}.
  \bibinfo{journal}{{\it J. Phys. Chem.\/}} \bibinfo{volume}{99}:\bibinfo{pages}{13118}

\bibitem{Garde:96:b}
\bibinfo{author}{\bibnamefont{Garde} \bibfnamefont{S}},
  \bibinfo{author}{\bibnamefont{Hummer} \bibfnamefont{G}},
  \bibinfo{author}{\bibnamefont{Paulaitis} \bibfnamefont{ME.}} \bibinfo{year}{1996}.
  \bibinfo{journal}{{\it Faraday Disc.\/}} \bibinfo{volume}{103}:\bibinfo{pages}{125}

\bibitem{AshbaughHS:Hydces}
\bibinfo{author}{\bibnamefont{Ashbaugh} \bibfnamefont{HS}},
  \bibinfo{author}{\bibnamefont{Kaler} \bibfnamefont{EW}},
  \bibinfo{author}{\bibnamefont{Paulaitis} \bibfnamefont{ME.}} \bibinfo{year}{1998}.
  \bibinfo{journal}{{\it Biophys. J.\/}} \bibinfo{volume}{75}:\bibinfo{pages}{755}
  
  
  
\bibitem{HummerG:Hydffa}
\bibinfo{author}{\bibnamefont{Hummer} \bibnamefont{G}}.  \bibinfo{year}{1999}.
  \bibinfo{journal}{{\it J. Am. Chem. Soc.\/}}
\bibinfo{volume}{121}:\bibinfo{pages}{6299--6305}

\bibitem{BorodinO:moldss}
\bibinfo{author}{\bibnamefont{Borodin} \bibnamefont{O,}}
\bibinfo{author}{\bibnamefont{Bedrov} \bibnamefont{D,}}
\bibinfo{author}{\bibnamefont{Smith} \bibnamefont{GD.}} \bibinfo{year}{2001}.
  \bibinfo{journal}{{\it Macromol.\/}}
\bibinfo{volume}{34}:\bibinfo{pages}{5687--5693}


\bibitem{AkasakaK:Lowesp}
\bibinfo{author}{\bibnamefont{Akasaka} \bibfnamefont{K,}}
  \bibinfo{author}{\bibnamefont{Li} \bibfnamefont{H.}} \bibinfo{year}{2001}.
  \bibinfo{journal}{{\it Biochem.\/}} \bibinfo{volume}{40}:\bibinfo{pages}{8665}
  
\bibitem{KAUZMANNW:PROSTU}
\bibinfo{author}{\bibnamefont{Kauzmann} \bibfnamefont{W.}} \bibinfo{year}{1987}.
  \bibinfo{journal}{{\it Nature\/}} \bibinfo{volume}{325}:\bibinfo{pages}{763--764}

\bibitem{wallqvista:premsa}
\bibinfo{author}{\bibnamefont{Wallqvist} \bibfnamefont{A.}} \bibinfo{year}{1992}.
  \bibinfo{journal}{{\it J. Chem. Phys.\/}} \bibinfo{volume}{96}:\bibinfo{pages}{1655}

\bibitem{Richards:SA:91}
\bibinfo{author}{\bibnamefont{Richards} \bibfnamefont{FM.}} \bibinfo{year}{1991}.
  \bibinfo{journal}{{\it Sci. Am.\/}} \bibinfo{volume}{264}:\bibinfo{pages}{54}


\bibitem{Wallqvist:JPC:95:b}
\bibinfo{author}{\bibnamefont{Wallqvist} \bibfnamefont{A,}}
  \bibinfo{author}{\bibnamefont{Berne} \bibfnamefont{BJ.}} \bibinfo{year}{1995}.
  \bibinfo{journal}{{\it J. Phys. Chem.\/}} \bibinfo{volume}{99}:\bibinfo{pages}{2893}
  
\bibitem{Lum:JPCB:99}
\bibinfo{author}{\bibnamefont{Lum} \bibfnamefont{K}},
  \bibinfo{author}{\bibnamefont{Chandler} \bibfnamefont{D}},
  \bibinfo{author}{\bibnamefont{Weeks} \bibfnamefont{JD.}}  \bibinfo{year}{1999}.
  \bibinfo{journal}{{\it J. Phys. Chem. B\/}} \bibinfo{volume}{103}:\bibinfo{pages}{4570}


\bibitem{HuangDM:Scahsf}
\bibinfo{author}{\bibnamefont{Huang} \bibfnamefont{DM}},
  \bibinfo{author}{\bibnamefont{Geissler}} \bibfnamefont{P},
  \bibinfo{author}{\bibnamefont{Chandler} \bibfnamefont{D.}} \bibinfo{year}{2001}.
  \bibinfo{journal}{{\it J. Phys. Chem. B\/}} \bibinfo{volume}{105}:\bibinfo{pages}{6704}

\bibitem{HuangDM:Temlsd}
\bibinfo{author}{\bibnamefont{Huang} \bibfnamefont{DM,}}
  \bibinfo{author}{\bibnamefont{Chandler} \bibfnamefont{D.}} \bibinfo{year}{2000}.
  \bibinfo{journal}{{\it Proc. Nat. Acad. Sci. USA\/}} \bibinfo{volume}{97}:\bibinfo{pages}{8324}

\bibitem{HuangDM:Cavfdt}
\bibinfo{author}{\bibnamefont{Huang} \bibfnamefont{DM,}}
  \bibinfo{author}{\bibnamefont{Chandler} \bibfnamefont{D.}} \bibinfo{year}{2000}.
  \bibinfo{journal}{{\it Phys. Rev. E\/}} \bibinfo{volume}{61}:\bibinfo{pages}{1501}
  
  

\bibitem{SunSX:Weidft}
\bibinfo{author}{\bibnamefont{Sun} \bibfnamefont{SX}}.  \bibinfo{year}{2001}.
 \bibinfo{journal}{{\it Phys. Rev. E\/}} \bibinfo{volume}{64}:\bibinfo{pages}{1512}


\bibitem{ReintenWolde:2001}
\bibinfo{author}{\bibnamefont{Rein~ten~Wolde} \bibfnamefont{P}},
  \bibinfo{author}{\bibnamefont{Sun} \bibfnamefont{SX}},
  \bibinfo{author}{\bibnamefont{Chandler} \bibfnamefont{D.}} \bibinfo{year}{2001}.
\emph{\bibinfo{title}{Model of a fluid
  at small and large length scales and the hydrophobic effect}},   
  \bibinfo{journal}{private communication}  

\bibitem{MullerEA:Molssh}
\bibinfo{author}{\bibnamefont{M{\"{u}}ller} \bibfnamefont{E,}}
  \bibinfo{author}{\bibnamefont{Gubbins} \bibfnamefont{K.}} \bibinfo{year}{1998}.
  \bibinfo{journal}{{\it Carbon\/}} \bibinfo{volume}{36}:\bibinfo{pages}{1433}

\bibitem{Huang:2001}
\bibinfo{author}{\bibnamefont{Huang} \bibfnamefont{DM,}}
  \bibinfo{author}{\bibnamefont{Chandler} \bibfnamefont{D.}} \bibinfo{year}{2001}.
 \emph{\bibinfo{title}{The hydrophobic effect
  and the influence of solute-solvent attractions}},  
  \bibinfo{journal}{private communication} 

\bibitem{BrovchenkoI:Gibesw}
\bibinfo{author}{\bibnamefont{Brovchenko} \bibfnamefont{I}},
  \bibinfo{author}{\bibnamefont{Paschek, \bibfnamefont{D}}},
  \bibinfo{author}{\bibnamefont{Geiger} \bibfnamefont{A}},
  \bibinfo{journal}{{\it J. Chem. Phys.\/}} \bibinfo{volume}{113}:\bibinfo{pages}{5026} 
  \bibinfo{year}{2000}.

\bibitem{WallqvistA:modsdh}
\bibinfo{author}{\bibnamefont{Wallqvist} \bibfnamefont{A}},
  \bibinfo{author}{\bibnamefont{Gallicchio} \bibfnamefont{E}},
  \bibinfo{author}{\bibnamefont{Levy} \bibfnamefont{RM.}~} \bibinfo{year}{2001}.
  \bibinfo{journal}{{\it J. Phys. Chem. B\/}} \bibinfo{volume}{105}:\bibinfo{pages}{6745}

\bibitem{BrovchenkoI:Phaewc}
\bibinfo{author}{\bibnamefont{Brovchenko} \bibfnamefont{I}},
  \bibinfo{author}{\bibnamefont{Geiger} \bibfnamefont{A}},
  \bibinfo{author}{\bibnamefont{Oleinikova} \bibfnamefont{A.}} \bibinfo{year}{2001}.
  \bibinfo{journal}{{\it Phys. Chem. Chem. Phys.\/}} \bibinfo{volume}{3}:\bibinfo{pages}{1567}

\bibitem{Hummer:2001}
\bibinfo{author}{\bibnamefont{Hummer}} \bibfnamefont{G},
  \bibinfo{author}{\bibnamefont{Rasaiah} \bibfnamefont{JC}},
  \bibinfo{author}{\bibnamefont{Noworyta} \bibfnamefont{J.}}  \bibinfo{year}{2001}.
\emph{\bibinfo{title}{Water conduction
  through a hydrophobic channel of a carbon nanotube}},  
  \bibinfo{journal}{ private communication} 

\end{thebibliography}
\end{document}